\documentclass[aps, prb, english, notitlepage, superscriptaddress, floatfix]{revtex4-2} 
\usepackage{bm, color}
\usepackage{placeins}
\usepackage{nameref}
\usepackage{amssymb}
\usepackage{graphicx}
\usepackage{subfigure}
\usepackage{dcolumn}
\usepackage{bm}
\usepackage{feynmf}
\usepackage{mathtools, nccmath}
\usepackage{amsthm}

\allowdisplaybreaks
\usepackage[colorlinks = true,
            allcolors  = blue]{hyperref}
\usepackage{color,soul}

\def\C{\mathbb{C}}
\def\G{\mathcal{G}}
\def\eC{\bar{\mathbb{C}}}

\usepackage{CJK}
\usepackage{hyperref}

\def\aver{\rangle_0^{K^{-1}}}
\def\ivk{(K^{-1})}

\preprint{APS/123-QED}
\begin{document}


\title{Strong-coupling expansion and two-point Pad\'e approximation for lattice $\phi^4$ field theory}

\author{Yuanran Zhu}%
\thanks{Corresponding author:yzhu4@lbl.gov}
\affiliation{Applied Mathematics and Computational Research Division, Lawrence Berkeley National Laboratory, Berkeley, USA, 94720.}

\author{Efekan K\"okc\"u}%
\affiliation{Department of Electrical and Computer Engineering, University of Central Florida, Orlando, FL, USA, 32816.}






\author{Chao Yang}%
\affiliation{Applied Mathematics and Computational Research Division, Lawrence Berkeley National Laboratory, Berkeley, USA, 94720.}






\begin{abstract}
Reliable approximations for correlation functions at intermediate and strong coupling remain hard to obtain for general quantum field theories. Perturbative expansions are often asymptotic or have a finite radius of convergence, which limits their applicability beyond weak coupling. Here we combine weak- and strong-coupling expansions and propose to use two-point Pad\'e schemes to construct approximants. For lattice $\phi^4$ theory, we show that this two-point interpolation strategy yields accurate global approximations to the two-point correlation function across broad coupling regimes and compares favorably with standard one-point resummation methods. We also provide heuristic explanations for the observed convergence behavior and discuss the practical range of validity of the approach.
\end{abstract}

\maketitle
\section{Introduction}
\label{sec:intro}
Perturbative expansions are basic calculation tools in statistical mechanics and quantum field theory. Their practical limitations are also well known: standard many-body perturbation theory typically produces a weak-coupling expansion in the interaction strength $g$, but this series often has a finite radius of convergence or is only asymptotic. Consequently, finite-order truncations can lose predictive power precisely in the intermediate- and strong-coupling regimes, where genuinely non-perturbative phenomena emerge.

For numerical investigations of intrinsically strongly interacting systems, such as quantum chromodynamics and strongly correlated materials, one typically pursues one of two broad strategies. The first is to develop fundamentally non-perturbative numerical methods, including lattice Monte Carlo \cite{blankenbecler1981monte,ceperley1995path,creutz1979monte} and tensor network techniques~\cite{white1992DMRG,schollwock2005density}, which can yield quite accurate results for specific systems but are often computationally demanding and may face additional obstacles such as the fermionic sign problem. The second retains a perturbative starting point but aims to extend the validity of perturbative series beyond their naive convergence domain. This route is attractive because it can reuse a large body of diagrammatic perturbation results and often offers favorable computational scalability. 
Motivated by these potential advantages, over the past decades, a variety of resummation and analytic continuation techniques have been developed to extend the convergence radius of perturbative expansions, with prominent examples including Pad\'e approximants~\cite{baker1981pade,baker1981continuous}, Borel resummation~\cite{parisi1988statistical}, and, more recently, resurgence-based approaches~\cite{dorigoni2019introduction,serone2018lambdaphi4}. While these methods can be remarkably successful in favorable cases, they typically rely on information from a single perturbation series (most often the weak-coupling expansion about 
$g=0$). As a consequence, the analytic continuation is inherently one-sided and the associated errors are difficult to control a priori, making it difficult to get highly accurate approximation results. This motivates constructing approximations that incorporate additional asymptotic information whenever it is available. The adopted method in this work, namely the two-point Pad\'e expansion, is a simple and practical way to achieve this goal. The main idea is quite direct: instead of relying solely on the weak-coupling expansion, we develop a strong-coupling expansion for the interacting system that is valid as $g\to +\infty$. Then, by building rational approximants that match both the weak-coupling expansion (WCE) about \( g\rightarrow 0 \) and the strong-coupling expansion (SCE) about \( g \to +\infty \), we can effectively {\em interpolate} between the two asymptotic regimes, yielding a natural global approximation for correlation functions that is generally valid for a wide range of interaction strengths.

We note that the idea of using two-point Pad\'e expansions to construct global approximations is not new, and has been explored in high-energy physics models, e.g., Refs.~\cite{sen2013s,banks2013two,honda2014perturbation}. This work will mainly focus on the lattice $\phi^4$ theory. In this context, comparing with the previous works our contributions are twofold. First, constructing a two-point Pad\'e expansion requires access to a strong-coupling expansion for the underlying field theory. In Refs.~\cite{sen2013s,banks2013two,honda2014perturbation}, the strong-coupling information is often obtained from the weak-coupling expansion via duality arguments. Here, by contrast, we derive the strong-coupling expansion for the lattice $\phi^4$ theory directly, using an approach tailored to yield explicit combinatorial formulas for the coefficients. This facilitates systematic, computer-aided generation of high-order expansions, while complementing the classical strong-coupling results of Bender \textit{et al.}~\cite{bender1979strong,bender1980strong,bender1981strong}. Second, we apply the resulting two-point Pad\'e expansion to the two-point correlation function of lattice $\phi^4$ theory, benchmark it against Pad\'e-type approximations built only from the WCE, and also compare with Borel resummation. Our numerical results show that the two-point Pad\'e expansion yields convergent approximations to the correlation function and has clear numerical advantages over one-point Pad\'e approximants based solely on the WCE. We further provide a heuristic explanation for the convergence behavior and discuss conditions under which numerical convergence can be expected more generally.

The remainder of the paper is organized as follows. In Sec.~\ref{sec:sce} we derive the strong-coupling expansion for lattice \(\phi^4\) theory and present its combinatorial/diagrammatic formulation. In Sec.~\ref{sec:2pade_intro} we construct two-point Pad\'e approximants, present numerical results and comparisons, and discuss convergence behavior and the reason behind it. The main conclusions of the paper are summarized in Sec.~\ref{sec:Conclusion}.

\section{Strong coupling expansion for lattice \(\phi^4\) field theory}
\label{sec:sce}
Consider the Hamiltonian for lattice \( \phi^4 \) field theory:
\begin{align}\label{lattice_phi_4}
\mathcal{H} = \sum_{ij} \frac{t_{ij}}{2}(\phi_i - \phi_j)^2 + \sum_i \frac{\mu}{2} \phi_i^2 + \frac{g}{4!} \phi_i^4,
\end{align}
where \( t_{ij} \) denotes the nearest-neighbor hopping matrix, 
subject to appropriate boundary conditions. $g>0$ is the coupling strength of the field, and \( \mu \) is the mass parameter.
This discrete form of the Hamiltonian is universal for \( \phi^4 \) fields in any spatial dimension. We seek a strong-coupling expansion for the two-point correlation function:
\begin{align}\label{two-point_G}
G_{ij} = \frac{\int D[\phi]\, \phi_i \phi_j\, e^{-\mathcal{H}[\phi]}}{\int D[\phi]\, e^{-\mathcal{H}[\phi]}},
\end{align}
valid in the limit \( g \to +\infty \). This strong coupling expansion for $\phi^4$ fields has
been studied thoroughly in the classical work by Bender et al.~\cite{bender1979strong,bender1980strong,bender1981strong}. 
In this section, we present an alternative derivation of the strong-coupling expansion (SCE) series, 
using an approach inspired by the work of Pairault et al.~\cite{pairault1998strong,pairault2000strong}.
Our goal is to obtain explicit combinatorial formulas for the expansion coefficients that are amenable to efficient numerical evaluation. The key steps are: (i) use a Hubbard--Stratonovich transform to turn the quadratic part into a Gaussian auxiliary-field integral, (ii) integrate out the original field site-by-site to obtain local moments and an expansion in powers of \( g^{-1/2} \), and (iii) evaluate the resulting Gaussian moments using Wick's theorem and organize them diagrammatically. 

To begin, we rewrite the partition function as: 
\begin{align*}
Z=\int D[\phi]e^{-\frac{1}{2}\langle\phi|K|\phi\rangle-\tilde S_0[\phi]},
\qquad
\tilde S_0[\phi]=\sum_{i=1}^N\frac{g}{4!}\phi_i^4
\end{align*}
where $\frac{1}{2}\langle\phi|K|\phi\rangle=\sum_{ij} t_{ij} (\phi_i - \phi_j)^2+\sum_i\frac{\mu}{2}\phi_i^2$. The Hubbard--Stratonovich (HS) transform for the Gaussian integral is given by:
\[
e^{-\frac{1}{2}\langle\phi|K|\phi\rangle}
=\frac{1}{\sqrt{(2\pi)^N\text{det}(K)}}
\int D[\psi]e^{-\frac{1}{2}\langle\psi|K^{-1}|\psi\rangle-i\langle\psi|\phi\rangle}
\]
where \( N \) is the total number of lattice sites. \( \psi \) is an auxiliary field with Gaussian weight \( \exp\{-\frac{1}{2}\langle \psi | K^{-1} | \psi \rangle\} \). In the following derivation, since the normalization factor in front of the integral transform appears in both the numerator and denominator, it will cancel out and is therefore omitted. After the HS transform, the partition function can be written as
\begin{align}\label{HS_Z}
Z &\propto 
\int D[\phi] D[\psi]\, e^{ -\frac{1}{2} \langle \psi | K^{-1} | \psi \rangle - i \langle \psi | \phi \rangle }\, e^{ -\tilde{S}_0[\phi] } 
= \int D[\psi]\, e^{ -\frac{1}{2} \langle \psi | K^{-1} | \psi \rangle } \left\langle e^{-i \langle \psi | \phi \rangle } \right\rangle_0^S,
\end{align}
where $\langle\cdot\rangle_0^S$ denotes the ensemble average with respect to $e^{-\tilde S_0[\phi]}$:
\[
\langle\cdot\rangle_0^S=\left\langle\int D[\phi](\cdot)e^{-\tilde S_0[\phi]}
\right\rangle
\]
After taking the average over the \( \phi \)-field in the mixed exponential term 
\( \left\langle e^{-i \langle \psi | \phi \rangle } \right\rangle_0^S \), 
only the \( \psi \)-field remains. 
Since the weight for \( \psi \), i.e. $e^{ -\frac{1}{2} \langle \psi | K^{-1} | \psi \rangle } $
 is Gaussian, one can restore Wick's theorem and apply 
 conventional diagrammatic techniques to obtain the perturbative expansion.
Let us first evaluate the ensemble average \( \left\langle e^{-i \langle \psi | \phi \rangle } \right\rangle_0^S \).
Due to the diagonality of \( \tilde S_0[\phi] \),
each \( \phi_i \) is independent, 
and we can factorize the average as:
\begin{align}\label{average_factorization}
\left\langle e^{-i\langle\psi|\phi\rangle}\right\rangle_0^S
=\left\langle e^{-i\sum_{j=1}^N\psi_j\phi_j}\right\rangle_0^S
=\prod_{j=1}^N\left\langle e^{-i\psi_j\phi_j}\right\rangle_0^s,
\end{align}
where \( \langle \cdot \rangle_0^s \) denotes the ensemble average with respect to the distribution 
\( \rho(\phi_i) \propto e^{-\frac{g}{4!} \phi_i^4} \), which is the same for each site \( i \). The even-order statistical moments for this distribution are known, with all odd moments vanishing:
\begin{align}\label{moments_phi_i}
\mu_{2n}(g)=\frac{1}{2}\left(\frac{g}{4!}\right)^{-\frac{2n+1}{4}} \Gamma\left(\frac{2n+1}{4}\right).
\end{align}
Thus, we obtain the expansion:
\begin{align}\label{phi_i_expansion}
\left\langle e^{-i\psi_i\phi_i}\right\rangle_0^s \propto \sum_{n=0}^{\infty} (-1)^n \frac{\psi_i^{2n}}{(2n)!} \mu_{2n}(g),
\end{align}
where, again, the normalization constant is omitted as it will cancel in the final expression. Substituting \eqref{phi_i_expansion} into \eqref{average_factorization}, and then using the generalized Cauchy product formula, we have:
\begin{equation}\label{mix_exp_expand}
\begin{aligned}
\left\langle e^{-i\langle\psi|\phi\rangle}\right\rangle_0^S
&=\prod_{i=1}^N\left\langle e^{-i\psi_i\phi_i}\right\rangle_0^s\\
&\propto
\left[\frac{1}{2}\sum_{j_1=0}^{\infty}\frac{(-1)^{j_1}\left(\frac{g}{4!}\right)^{-\frac{2j_1+1}{4}}}{(2j_1)!}\psi_1^{2j_1}\Gamma\left(\frac{2j_1+1}{4}\right)\right]
\cdots
\left[\frac{1}{2}\sum_{j_N=0}^{\infty}\frac{(-1)^{j_N}\left(\frac{g}{4!}\right)^{-\frac{2j_N+1}{4}}}{(2j_N)!}\psi_N^{2j_N}\Gamma\left(\frac{2j_N+1}{4}\right)\right]\\
&=\sum_{m=0}^{\infty}(-1)^{m}\left(\frac{g}{4!}\right)^{-\frac{2m+1}{4}}\sum_{\substack{j_1 + \cdots + j_N = m \\ j_i \geq 0}}
\frac{1}{2^N}\frac{\psi_1^{2j_1}\cdots\psi_N^{2j_N}}{(2j_1)!(2j_2)!\cdots(2j_N)!}\Gamma\left(\frac{2j_1+1}{4}\right)\cdots\Gamma\left(\frac{2j_N+1}{4}\right)\\
&=\sum_{m=0}^{\infty}(-1)^mg^{-\frac{2m+1}{4}}\sum_{\substack{j_1 + \cdots + j_N = m \\ j_i \geq 0}}
W(j_1,\cdots,j_N)\psi_1^{2j_1}\cdots\psi_N^{2j_N}
\end{aligned}
\end{equation}
where $W(j_1,\cdots,j_N)$ is a scalar function, defined by:
\[
W(j_1,\cdots,j_N)
=\frac{\left(\frac{1}{4!}\right)^{-\frac{2m+1}{4}}}{2^N}\prod_{i=1}^N\frac{1}{(2j_{i})!}\Gamma\left(\frac{2j_i+1}{4}\right)
\]
From the definition, we observe that \( W(j_1, \ldots, j_N) \)
is a symmetric function; that is, 
permuting the indices \( j_1, \ldots, j_N \) does not change 
the value of the function. For this reason, in the following derivations,
we adopt a shorthand notation: for example, \( W(\bm{0}) = W(0, \ldots, 0) \), 
and \( W(2,1,1,\bm{0}) = W(2,1,1,0,\ldots,0) \), where we explicitly list only the nonzero indices. 
On the other hand, the two-point correlation function can be obtained
 using the Schwinger approach 
 by taking functional derivatives with respect to the source field.
  To this end, we introduce the generating functional \( Z(J) \), 
  defined as~\cite{zinn2021quantum}:
\begin{align*}
Z[J]=\frac{1}{Z}\int D[\phi]e^{-\frac{1}{2}\langle\phi|K|\phi\rangle-\tilde S_0[\phi]+\langle J|\phi\rangle}
\end{align*}
Following the same procedure as for the partition function, after the HS transformation, we can rewrite the generating functional as
\begin{align*}
Z[J]&=\frac{1}{Z}\int D[\psi]e^{-\frac{1}{2}\langle\psi|K^{-1}|\psi\rangle}\left\langle e^{-\langle i\psi-J|\phi\rangle}\right\rangle_0^S
\end{align*}
By definition, the two-point correlation function $G_{ij}$ is given by:
\begin{align}\label{G_strong_def}
G_{ij}=\frac{\delta^2}{\delta J_i\delta J_j}Z[J]\biggr|_{J=0}&=\frac{1}{Z}\int D[\psi]e^{-\frac{1}{2}\langle\psi|K^{-1}|\psi\rangle}\frac{\delta^2}{\delta J_i\delta J_j}\left[\left\langle e^{-\langle i\psi-J|\phi\rangle}\right\rangle_0^S\right]\Biggr|_{J=0}
\end{align}
Noticing that:
\[
\frac{\delta^2}{\delta J_i\delta J_j}\left[\left\langle e^{-\langle i\psi-J|\phi\rangle}\right\rangle_0^S\right]\Biggr|_{J=0}
=-\frac{\delta^2\langle e^{-i\langle\psi|\phi\rangle}\rangle_0^S}{\delta \psi_i\delta \psi_j}
\]
Using integration by parts twice in \eqref{G_strong_def} and then using \eqref{mix_exp_expand}, we obtain:
\begin{equation}\label{G_SCE}
\begin{aligned}
G_{ij} &= \frac{1}{Z} \int D[\psi] \frac{\delta^2}{\delta \psi_i \delta \psi_j} e^{-\frac{1}{2} \langle \psi | K^{-1} | \psi \rangle} \left\langle e^{-i\langle \psi | \phi \rangle} \right\rangle_0^S \\
&= \frac{\displaystyle \sum_{m=0}^{\infty} (-1)^{m+1}g^{-\frac{2m+1}{4}} \sum_{\substack{j_1 + \cdots + j_N = m \\ j_i \geq 0}} W(j_1, \cdots, j_N) \int D[\psi] \frac{\delta^2}{\delta \psi_i \delta \psi_j} e^{-\frac{1}{2} \langle \psi | K^{-1} | \psi \rangle} \psi_1^{2j_1} \cdots \psi_N^{2j_N}} 
{\displaystyle \sum_{m=0}^{\infty} (-1)^mg^{-\frac{2m+1}{4}} \sum_{\substack{j_1 + \cdots + j_N = m \\ j_i \geq 0}} W(j_1, \cdots, j_N) \int D[\psi] e^{-\frac{1}{2} \langle \psi | K^{-1} | \psi \rangle} \psi_1^{2j_1} \cdots \psi_N^{2j_N}}
\end{aligned}
\end{equation}
After taking the second-order functional derivative, the numerator becomes
\begin{align*}
\text{Numerator}&=
\ivk_{ij}+\sum_{k_1k_2}\ivk_{ik_1}\ivk_{jk_2}\sum_{m=0}^{\infty} (-1)^{m+1}g^{-\frac{2m+1}{4}}\\
&\qquad\qquad\times \sum_{\substack{j_1 + \cdots + j_N = m \\ j_i \geq 0}}W(j_1,\cdots,j_N)\int D[\psi]\psi_{k_1}\psi_{k_2}e^{-\frac{1}{2}\langle\psi|K^{-1}|\psi\rangle}\psi_1^{2j_1}\cdots\psi_N^{2j_N}.
\end{align*}
Thus, Eq.~\eqref{G_SCE} can be rewritten as
\begin{equation}\label{numer_by_denom}
\begin{aligned}
G_{ij} 
&= \ivk_{ij}
+\sum_{k_1k_2}\ivk_{ik_1}\ivk_{jk_2}
\frac{\sum_{m=0}^{+\infty} (-1)^{m+1} g^{-\frac{2m+1}{4}}\bm D^{(m)}}
{\sum_{m=0}^{+\infty} (-1)^{m}g^{-\frac{2m+1}{4}}N^{(m)}}\\
&
=[{\bm G}_s^{(0)} + g^{-1/2}{\bm G}_s^{(1)} + g^{-1}{\bm G}_s^{(2)}+\cdots]_{ij}
\end{aligned}
\end{equation}
where 
\begin{equation}\label{D_m_N_m}
\begin{aligned}
\bm D^{(m)} &=\sum_{\substack{j_1 + \cdots + j_N = m \\ j_i \geq 0}}W(j_1,\cdots,j_N)\int D[\psi]\psi_{k_1}\psi_{k_2}e^{-\frac{1}{2}\langle\psi|K^{-1}|\psi\rangle}\psi_1^{2j_1}\cdots\psi_N^{2j_N}
\\
N^{(m)} &=\sum_{\substack{j_1 + \cdots + j_N = m \\ j_i \geq 0}}W(j_1,\cdots,j_N)\int D[\psi]e^{-\frac{1}{2}\langle\psi|K^{-1}|\psi\rangle}\psi_1^{2j_1}\cdots\psi_N^{2j_N}
\end{aligned}
\end{equation}
We see that Eq.~\eqref{numer_by_denom} takes the desired form of SCE series. Define the intermediate quantities \( \tilde{\bm{G}}_s^{(i)} \) via
\begin{align}\label{tilde_G_to_G}
\bm G_s^{(0)}= 
\begin{cases}
(K^{-1})_{ij}+\sum_{k_1k_2}\ivk_{ik_1}\ivk_{jk_2}\tilde{\bm G}_s^{(0)},\quad i=0\\
\sum_{k_1k_2}\ivk_{ik_1}\ivk_{jk_2}\tilde{\bm G}_s^{(i)},\quad i\geq 1
\end{cases}
\end{align}
So the SCE coefficients \( \bm{G}_s^{(0)}, \bm{G}_s^{(1)}, \bm{G}_s^{(2)}, \ldots \) are obtained from \( \tilde{\bm{G}}_s^{(0)}, \tilde{\bm{G}}_s^{(1)}, \tilde{\bm{G}}_s^{(2)}, \ldots \), which are determined by matching:
\begin{align}\label{matching_cond}
\frac{\sum_{m=0}^{+\infty} (-1)^{m+1} g^{-\frac{2m+1}{4}}\bm D^{(m)}}
{\sum_{m=0}^{+\infty} (-1)^{m}g^{-\frac{2m+1}{4}} N^{(m)}}
=\tilde{\bm G}_s^{(0)} + g^{-1/2}\tilde{\bm G}_s^{(1)} + g^{-1}\tilde{\bm G}_s^{(2)}+\cdots
\end{align}
The coefficients \( \tilde{\bm{G}}_s^{(n)} \) can be expressed diagrammatically in terms of connected Feynman diagrams; their evaluation involves a combinatorial reorganization of Gaussian moments (Feynman strings and Möbius inversion), which we detail in the next two subsections. For convenience, we first state the resulting SCE coefficients \( \bm{G}_s^{(n)} \) up to third order:
\begin{equation}\label{phi_4_SCE_firstfew}
\begin{aligned}
[\bm{G}_s^{(0)}]_{ij} &= 0, \\
[\bm{G}_s^{(1)}]_{ij} &= \frac{2 W(1,\bm{0})}{W(\bm{0})} I_{ij}, \\
[\bm{G}_s^{(2)}]_{ij} &= 
\left( -12 \frac{W(2,\bm{0})}{W(\bm{0})} + 6 \frac{W(1,1,\bm{0})}{W(\bm{0})} \right) K_{ii} \delta_{ij} 
- 4 \frac{W(1,1,\bm{0})}{W(\bm{0})} K_{ij}, \\
[\bm{G}_s^{(3)}]_{ij} &= 
\left( 90 \frac{W(3,\bm{0})}{W(\bm{0})} - 90 \frac{W(2,1,\bm{0})}{W(\bm{0})} + 30 \frac{W(1,1,1,\bm{0})}{W(\bm{0})} \right) K_{ii} \delta_{ij} \\
&\quad + \left( 24 \frac{W(2,1,\bm{0})}{W(\bm{0})} - 12 \frac{W(1,1,1,\bm{0})}{W(\bm{0})} \right) 
\left[ K_{ii} K_{ij} + K_{ij} K_{jj} + \delta_{ij} \sum_k K_{ik} K_{ki} \right] \\
&\quad + 8 \frac{W(1,1,1,\bm{0})}{W(\bm{0})} \sum_k K_{ik} K_{kj}.
\end{aligned}
\end{equation}
The above SCE coefficients are consistent with the classical strong-coupling expansion obtained by Bender et al.~\cite{bender1979strong,bender1980strong,bender1981strong}.
In particular, by replacing their notation \( A_{2n} \) with the local moments \( \mu_{2n} \) defined in Eq.~\eqref{moments_phi_i} in Eq.~(2.10) of Ref.~\cite{bender1980strong}, one recovers the results above.

\subsection{Enumeration of Feynman diagrams}
In this section, we determine \( \tilde{\bm{G}}_s^{(n)} \). This reduces to evaluating the ensemble averages of \( \bm{D}^{(m)} \) and \( N^{(m)} \) in Eq.~\eqref{D_m_N_m}. Let us first focus on the denominator term \( N^{(m)} \). After taking the ensemble average over \( \psi_i \), the summation becomes:
\begin{align}\label{average_sum}
\sum_{\substack{j_1 + \cdots + j_N = m \\ j_i \geq 0}} W(j_1,\cdots,j_N) \left\langle \psi_1^{2j_1} \cdots \psi_N^{2j_N} \right\rangle_0^{K^{-1}},
\end{align}
where \( \langle \cdot \rangle_0^{K^{-1}} \) denotes the average with respect to the Gaussian weight \( e^{-\frac{1}{2} \langle \psi | K^{-1} | \psi \rangle} \). For the first step, we formally rewrite the permutation sum into a combination sum. We use $\{j_1,\cdots, j_N\}$ to represent the index set such that $j_1+\cdots+j_N =m, j_i\geq 0$. For instance, when $N=5, m=3$, there are $3$ possible index sets (up to permutation): $\{3,\bm 0\}$, $\{2,1,\bm 0\}$, and $\{1,1,1,\bm 0\}$. Since $W(j_1,\cdots,j_N)$ is a symmetric function, we can rewrite \eqref{average_sum} as:
\begin{align}\label{combi_sum}
\sum_{\substack{j_1 + \cdots + j_N = m \\ j_i \geq 0}}W(j_1,\cdots,j_N)\langle\psi_1^{2j_1}\cdots\psi_N^{2j_N}\rangle_0^{K^{-1}}
=\sum_{\substack{\{j_1,\cdots, j_N\},\\\text{where }j_1 + \cdots + j_N = m \\ j_i \geq 0}}W(j_1,\cdots j_N)F_{\{j_1,\cdots,j_N\}}
\end{align}

In Eq.~\eqref{combi_sum}, the first summation runs over all {\em permutations} of nonnegative indices \( \{j_1,\cdots,j_N\} \) such that \( j_1+\cdots+j_N = m \), while the second summation runs over all {\em combinations} of the index set. Thus, the formal expression \( F_{\{j_1,\cdots,j_N\}} \) collects all statistical moments \( \langle \cdots \rangle_0^{K^{-1}} \) that correspond to the same index set. We refer to \( F_{\{j_1,\cdots,j_N\}} \) as a \emph{Feynman string}, as we will show that it corresponds to a string of Feynman diagrams. Its subscript \( \{j_1,\cdots,j_N\} \) indicates that this quantity depends only on the index set itself. To explicitly construct \( F_{\{j_1,\cdots,j_N\}} \), we introduce the following definitions.

For a given index set \( \{j_1,\cdots, j_N\} \), suppose there are \( L \) non-zero elements. Among them, there are \( N_p \) repeated values, with repetition lengths \( l_i \) for \( 1 \leq i \leq N_p \). For example, for the index set \( \{3,2,2,1,1,1,0,0\} \), we have \( L = 6 \), \( N_p = 2 \), \( l_1 = 2 \), and \( l_2 = 3 \). For the non-zero elements, define the ordered set \( \{j^o_1,\cdots,j^o_L\} \) such that \( j^o_1 \geq \cdots \geq j^o_L \). In the example above, this ordered set is \( \{3,2,2,1,1,1\} \). With these definitions, the Feynman strings can be explicitly written as:
\begin{align}\label{feynman_string}
F_{\{j_1,\cdots,j_N\}}
= \frac{1}{\prod_{i=1}^{N_p} l_i!} 
\sum_{k_1 \neq k_2 \neq \cdots \neq k_L}
\left\langle \psi_{k_1}^{2j^o_1} \psi_{k_2}^{2j^o_2} \cdots \psi_{k_L}^{2j^o_L} \right\rangle_0^{K^{-1}}.
\end{align}
One can verify that the number of statistical moments in the summation above is 
\(
\frac{N!}{(N - L)! \prod_{i=1}^{N_p} l_i!},
\)
which matches the total number of permutations associated with the index set \( \{j_1,\cdots,j_N\} \). The next step is to simplify the {\em exclusion sum} \( \sum_{k_1 \neq k_2 \neq \cdots \neq k_L}(\cdots) \). Although Wick's theorem still applies to this form, it cannot be directly evaluated using diagrammatics since the Feynman rules for vertex summation requires {\
{\em inclusion sums} such as \( \sum_{k_1k_2} \) or \( \sum_{k_1k_2k_3} \). Computationally, this leads to the following difficulties: unlike Feynman diagrams, exclusion sums cannot be directly rewritten as tensor contractions, making them unsuitable for efficient numerical implementation. Furthermore, the cancellation of disconnected diagrams is not manifest in this formulation. This technical difficulty can be resolved using the {\em Möbius inversion} formula over the poset of partitions of \( n \), ordered by refinement~\cite{stanley2011enumerative}. The Möbius inversion allows one to express exclusion sums as linear combinations of inclusion sums, with combinatorial coefficients. The general formula that works for any exclusion sum and noncommutative summand is given in Appendix \ref{sec:app1_mobius}. To obtain the SCE for $\phi^4$ field up to third order, we only need the following result:
\begin{equation}\label{mobius_inversion_example}
\begin{aligned}
\sum_{k_1\neq k_2}f(k_1,k_2)
=&\sum_{k_1k_2}f(k_1,k_2) - \sum_{k_1}f(k_1,k_1)\\
\sum_{k_1\neq k_2\neq k_3}f(k_1,k_2,k_3)
=&\sum_{k_1k_2k_3}f(k_1,k_2,k_3)\\
-&\sum_{k_1k_2}f(k_1,k_1,k_2)
-\sum_{k_1k_2}f(k_1,k_2,k_1)
-\sum_{k_1k_2}f(k_1,k_2,k_2)\\
+&2\sum_{k_1}f(k_1,k_1,k_1)
\end{aligned}
\end{equation}



For our case, the arguments of $f$ are the site indices $k_1,\cdots, k_L$ in the Feynman string \eqref{feynman_string}. As an example, applying this formula for index set $\{1,1,1,0,0\}$, the corresponding Feynman string can be decomposed as:
\begin{align*}
F_{\{1,1,1,0,0\}}&= \frac{1}{3!}\sum_{k_1\neq k_2\neq k_3}^N\langle\psi_{k_1}^2\psi_{k_2}^2\psi_{k_3}^2\rangle_0^{K^{-1}}\\
&=\frac{1}{3!}
\left[\sum_{k_1 k_2k_3}^N\langle\psi_{k_1}^2\psi_{k_2}^2\psi_{k_3}^2\rangle_0^{K^{-1}}
-3\sum_{k_1k_2}^N\langle\psi_{k_1}^4\psi_{k_2}^2\rangle_0^{K^{-1}}
+2\sum_{k_1}^N\langle\psi_{k_1}^6\rangle_0^{K^{-1}}
\right]
\end{align*}
As an example for non-symmetric case: for index set $\{3,1,1,0,0\}$, the corresponding Feynman string can be decomposed as:
\begin{align*}
F_{\{3,1,1,0,0\}}&= \frac{1}{2!2!}\sum_{k_1\neq k_2\neq k_3}^N\langle\psi_{k_1}^6\psi_{k_2}^2\psi_{k_3}^2\rangle_0^{K^{-1}}\\
&=\frac{1}{2!2!}
\left[\sum_{k_1 k_2k_3}^N\langle\psi_{k_1}^6\psi_{k_2}^2\psi_{k_3}^2\rangle_0^{K^{-1}}
-\sum_{k_1k_2}^N\langle\psi_{k_1}^6\psi_{k_2}^4\rangle_0^{K^{-1}}
-2\sum_{k_1k_2}^N\langle\psi_{k_1}^2\psi_{k_2}^8\rangle_0^{K^{-1}}
+2\sum_{k_1}^N\langle\psi_{k_1}^6\rangle_0^{K^{-1}}
\right]
\end{align*}
Combining the series expansion \eqref{numer_by_denom}-\eqref{D_m_N_m}, the definition of Feynman strings \eqref{feynman_string}, and the Möbius inversion formula \eqref{mobius_inversion_example}, we can obtain the SCE series and its diagrammatic representations. 
%
%
\subsection{Strong coupling expansion and diagrammatic representation}

Using the prescribed recipe,  by substituting Eqs.~\eqref{combi_sum}, \eqref{feynman_string}, and \eqref{mobius_inversion_example} into Eq.~\eqref{D_m_N_m}, and subsequently expanding the numerator of the fraction in Eq.~\eqref{numer_by_denom} up to third order, we obtain:
\begin{equation}\label{phi_4_numerator}
\begin{aligned}
\text{Numerator}= 
&-g^{-\frac{1}{4}}W(\bm 0)\langle \psi_{k_1}\psi_{k_2}\aver\\
&+g^{-\frac{3}{4}}W(1,\bm 0)\sum_{k_3}\langle \psi_{k_1}\psi_{k_2}\psi_{k_3}^2\aver\\
&-g^{-\frac{5}{4}}\left\{\left[W(2,\bm 0)-\frac{1}{2}W(1,1,\bm 0)\right]\sum_{k_3}\langle \psi_{k_1}\psi_{k_2}\psi_{k_3}^4\aver
+\frac{1}{2}W(1,1,\bm 0)
\sum_{k_3k_4}\langle\psi_{k_1}\psi_{k_2}\psi_{k_3}^2\psi_{k_3}^2\aver
\right\}\\
&+g^{-\frac{7}{4}}\Bigg\{\left[W(3,\bm 0)-W(2,1,\bm 0)+\frac{1}{3}W(1,1,\bm 0)\right]\sum_{k_3}\langle \psi_{k_1}\psi_{k_2}\psi_{k_3}^6\aver\\
&\qquad\quad
+
\left[W(2,1,\bm 0)-\frac{1}{2}W(1,1,1,\bm 0)\right]
\sum_{k_3k_4}\langle\psi_{k_1}\psi_{k_2}\psi_{k_3}^2\psi_{k_4}^2\aver\\
&\qquad\quad
+\frac{1}{6}W(1,1,1,\bm 0)
\sum_{k_3k_4k_5}\langle\psi_{k_1}\psi_{k_2}\psi_{k_3}^2\psi_{k_4}^2\psi_{k_5}^2\aver
\Bigg\}\\
&+\cdots
\end{aligned}
\end{equation}
Hereafter, we always consider the general case \( N \geq 3 \), ensuring that terms such as \( W(1,1,1,\boldsymbol{0}) \) does not vanish. For \( N < 3 \), the derivation is analogous and straightforward. For the denominator part, similarly we have: 
\begin{equation}\label{phi_4_denominator}
\begin{aligned}
\text{Denominator}= 
&-g^{-\frac{1}{4}}W(\bm 0)\\
&+g^{-\frac{3}{4}}W(1,\bm 0)\sum_{k_3}\langle \psi_{k_3}^2\aver\\
&-g^{-\frac{5}{4}}\left\{\left[W(2,\bm 0)-\frac{1}{2}W(1,1,\bm 0)\right]\sum_{k_3}\langle \psi_{k_3}^4\aver
+\frac{1}{2}W(1,1,\bm 0)
\sum_{k_3k_4}\langle\psi_{k_3}^2\psi_{k_3}^2\aver
\right\}\\
&+g^{-\frac{7}{4}}\Bigg\{\left[W(3,\bm 0)-W(2,1,\bm 0)+\frac{1}{3}W(1,1,\bm 0)\right]\sum_{k_3}\langle \psi_{k_3}^6\aver\\
&\qquad\quad
+
\left[W(2,1,\bm 0)-\frac{1}{2}W(1,1,1,\bm 0)\right]
\sum_{k_3k_4}\langle\psi_{k_3}^2\psi_{k_4}^2\aver\\
&\qquad\quad
+\frac{1}{6}W(1,1,1,\bm 0)
\sum_{k_3k_4k_5}\langle\psi_{k_3}^2\psi_{k_4}^2\psi_{k_5}^2\aver
\Bigg\}\\
&+\cdots
\end{aligned}
\end{equation}
Applying Wick's theorem to each term, and then using the matching condition ~\eqref{matching_cond}, we can derive $\tilde{\bm G}_s^{(n)}$. The whole process can be made simpler using Feynman diagrams which translates into tensor contractions with the following rules:

\begin{enumerate}
\item Each solid line
\begin{fmffile}{FU_01}
\raisebox{-0.15cm}{
    \resizebox{0.03\textwidth}{!}{ 
    \begin{fmfgraph*}(20,20)
      \fmfleft{i}
      \fmfright{j}
      \fmfv{d.shape=circle, d.filled=full, d.size=1mm}{i}  %
      \fmfv{d.shape=circle, d.filled=full, d.size=1mm}{j}  %
      \fmf{plain}{i,j}
    \end{fmfgraph*}
    } 
}
\end{fmffile}
connecting $i,j$ corresponds to matrix $K_{ij}$; Dashed line 
\begin{fmffile}{FU_02}
\raisebox{-0.15cm}{
    \resizebox{0.03\textwidth}{!}{ 
    \begin{fmfgraph*}(20,20)
      \fmfleft{i}
      \fmfright{j}
      \fmfv{d.shape=circle, d.filled=full, d.size=1mm}{i}  %
      \fmfv{d.shape=circle, d.filled=full, d.size=1mm}{j}  %
      \fmf{dashes}{i,j}
    \end{fmfgraph*}
    } 
}
\end{fmffile} corresponds to matrix $K^{-1}_{ij}$.
\item Assign each vertex a summation over the vertex point.
\item Assign each diagram a symmetry factor which equals to the number of permutations for all lines and vertices. 
\end{enumerate}
Multiplying the denominator of Eq.~\eqref{matching_cond} to the right-hand side and matching the series to order \( O(g^{-1/4}) \), we find:
\[
[\tilde{\bm G}_s^{(0)}]_{k_1k_2} = -\langle \psi_{k_1} \psi_{k_2} \rangle = -K_{k_1k_2}.
\]
where the diagrammatic representation simply is: 
$-$\begin{fmffile}{DC_g14_01}
\raisebox{-0.15cm}{
    \resizebox{0.03\textwidth}{!}{ 
    \begin{fmfgraph*}(20,20)
      \fmfleft{i}
      \fmfright{j}
      \fmfv{d.shape=circle, d.filled=full, d.size=1mm}{i}  %
      \fmfv{d.shape=circle, d.filled=full, d.size=1mm}{j}  %
      \fmf{plain}{i,j}
    \end{fmfgraph*}
    } 
}
\end{fmffile}
The result by matching the next order \( O(g^{-3/4}) \) is also straightforward to obtain:
\[
[\tilde{\bm G}_s^{(1)}]_{k_1k_2} = \frac{2W(1,\bm 0)}{W(\bm 0)} \sum_{k_3} \langle \psi_{k_1} \psi_{k_3} \rangle \langle \psi_{k_3} \psi_{k_2} \rangle =  \frac{2W(1,\bm 0)}{W(\bm 0)}\sum_{k_3} K_{k_1k_3} K_{k_3k_2}.
\]
where the diagrammatic representation is: 
$ \frac{2W(1,\bm 0)}{W(\bm 0)}$\begin{fmffile}{DC_g34_01}
\raisebox{-0.15cm}{
    \resizebox{0.04\textwidth}{!}{ 
    \begin{fmfgraph*}(20,20)
      \fmfleft{i}
      \fmfright{j}
      \fmfv{d.shape=circle, d.filled=full, d.size=1mm}{i}  %
      \fmfv{d.shape=circle, d.filled=full, d.size=1mm}{j}  %
      \fmfv{d.shape=circle, d.filled=full, d.size=1mm}{k1}  %
      \fmf{plain}{i,k1}
      \fmf{plain}{k1,j}
    \end{fmfgraph*}
    } 
}
\end{fmffile}
At the order $O(g^{-5/4})$, we have
\begin{align*}
\text{LHS}
&=\left[-W(2,\bm 0)+\frac{1}{2}W(1,1,\bm 0)\right]\sum_{k_3}\langle \psi_{k_1}\psi_{k_2}\psi_{k_3}^4\aver
-\frac{1}{2}W(1,1,\bm 0)
\sum_{k_3k_4}\langle\psi_{k_1}\psi_{k_2}\psi_{k_3}^2\psi_{k_3}^2\aver\\
\text{RHS}&=
W(\bm 0)[\tilde{\bm G}_s^{(2)}]_{k_1k_2}
-W(1,\bm 0)[\tilde{\bm G}_s^{(1)}]_{k_1k_2}\sum_{k_4}\langle \psi_{k_4}^2\aver\\
&\quad+\left[-W(2,\bm 0)+\frac{1}{2}W(1,1,\bm 0)\right][\tilde{\bm G}_s^{(1)}]_{k_1k_2}\sum_{k_3}\langle \psi_{k_3}^4\aver
-\frac{1}{2}W(1,1,\bm 0)
[\tilde{\bm G}_s^{(1)}]_{k_1k_2}\sum_{k_3k_4}\langle\psi_{k_3}^2\psi_{k_3}^2\aver
\end{align*}
Using Wick's theorem to factorize all the terms on the LHS, we first observe that the last two terms on the RHS will cancel out all the disconnected diagrams in the last two terms on the LHS in which \( k_1 \) and \( k_2 \) are connected. Moreover, since 
\[
W(1,1,\bm 0)=\frac{[W(1,\bm 0)]^2}{W(\bm 0)}=2\left(\frac{1}{4!}\right)^{-\frac{5}{4}}\frac{\left[\Gamma\left(\frac{3}{4}\right)\right]^{2}\left[\Gamma\left(\frac{1}{4}\right)\right]^{N-2}}{\left[\Gamma\left(\frac{1}{4}\right)\right]^{N}}
=2\left(\frac{1}{4!}\right)^{-\frac{5}{4}}\frac{\left[\Gamma\left(\frac{3}{4}\right)\right]^{2}}{\left[\Gamma\left(\frac{1}{4}\right)\right]^{2}},
\]
the disconnected Feynman diagram 
\begin{fmffile}{DC_g54_02}
\raisebox{-0.2cm}{
    \resizebox{0.04\textwidth}{!}{ 
    \begin{fmfgraph*}(20,20)
      \fmfleft{i}
      \fmfright{j}
      \fmfv{d.shape=circle, d.filled=full, d.size=1mm}{i}  %
      \fmfv{d.shape=circle, d.filled=full, d.size=1mm}{j}  %
      \fmfv{d.shape=circle, d.filled=full, d.size=1mm}{k1}  %
      \fmf{plain}{i,k1}
      \fmf{plain}{k1,j}
    \end{fmfgraph*}
    } 
}
\end{fmffile}
\begin{fmffile}{DC_g54_01}
\raisebox{-0.1cm}{
    \resizebox{0.04\textwidth}{!}{ 
    \begin{fmfgraph*}(20,20)
      \fmftop{i}
      \fmfbottom{j}
      \fmfv{d.shape=circle, d.filled=full, d.size=1mm}{j}  %
      \fmf{plain,left,tension=0.5}{i,j}
      \fmf{plain,left,tension=0.5}{j,i}
    \end{fmfgraph*}
    } 
}
\end{fmffile}
which correspond to the $[\tilde{\bm G}_s^{(1)}]_{k_1k_2}\sum_{k_4}\langle \psi_{k_4}^2\aver$ term on the RHS, cancels with the equivalent disconnected diagram arising from the last term on the LHS, owing to a symmetry factor of 4. Consequently, only the connected diagrams survive on the LHS, yielding
\begin{align*}
\tilde{\bm G}_s^{(2)}
=
\left[-12\frac{W(2,\bm 0)}{W(\bm 0)}+6\frac{W(1,1,\bm 0)}{W(\bm 0)}\right]
\begin{fmffile}{DC_G54_1}
\raisebox{-0.3cm}{
    \resizebox{0.05\textwidth}{!}{ 
    \begin{fmfgraph*}(20,20)
      \fmfleft{i}
      \fmfright{j}
      \fmfv{d.shape=circle, d.filled=full, d.size=1mm}{i}  %
      \fmfv{d.shape=circle, d.filled=full, d.size=1mm}{j}  %
      \fmfv{d.shape=circle, d.filled=full, d.size=1mm}{k1}  %
      \fmf{plain,tension=0.5}{k1,k1}
      \fmf{plain}{i,k1}
      \fmf{plain}{k1,j}
    \end{fmfgraph*}
    } 
}
\end{fmffile}
-4\frac{W(1,1,1)}{W(\bm 0)}
\begin{fmffile}{DC_G54_2}
\raisebox{-0.3cm}{
    \resizebox{0.05\textwidth}{!}{ 
    \begin{fmfgraph*}(20,20)
      \fmfleft{i}
      \fmfright{j}
      \fmfv{d.shape=circle, d.filled=full, d.size=1mm}{i}  %
      \fmfv{d.shape=circle, d.filled=full, d.size=1mm}{j}  %
      \fmfv{d.shape=circle, d.filled=full, d.size=1mm}{k1}  %
      \fmfv{d.shape=circle, d.filled=full, d.size=1mm}{k2}  %
      \fmf{plain}{i,k1}
      \fmf{plain}{k1,k2}
      \fmf{plain}{k2,j}
    \end{fmfgraph*}
    } 
}
\end{fmffile}
\end{align*}
Similarly, through some tedious calculation, we verified that all the disconnected diagrams at the order $O(g^{-7/4})$ also cancels out, which leads to:
\begin{align*}
\tilde{\bm G}_s^{(3)}
=&
\left[90\frac{W(3,\bm 0)}{W(\bm 0)}-90\frac{W(2,1,\bm 0)}{W(\bm 0)}+30\frac{W(1,1,1,\bm 0)}{W(\bm 0)}\right]
\begin{fmffile}{DC_G74_1}
\raisebox{-0.25cm}{
    \resizebox{0.05\textwidth}{!}{ 
    \begin{fmfgraph*}(20,20)
      \fmfleft{i}
      \fmfright{j}
      \fmfv{d.shape=circle, d.filled=full, d.size=1mm}{i}  %
      \fmfv{d.shape=circle, d.filled=full, d.size=1mm}{j}  %
      \fmfv{d.shape=circle, d.filled=full, d.size=1mm}{k1}  %
  \fmf{plain,left=0.6,tension=0.4}{k1,k1}
  \fmf{plain,right=0.6,tension=0.4}{k1,k1}
      \fmf{plain}{i,k1}
      \fmf{plain}{k1,j}
    \end{fmfgraph*}
    } 
}
\end{fmffile}
\\
&+\left[24\frac{W(2,1,\bm 0)}{W(\bm 0)}-12\frac{W(1,1,1,\bm 0)}{W(\bm 0)}\right]
\Bigg\{
\begin{fmffile}{DC_G74_2}
\raisebox{-0.3cm}{
    \resizebox{0.05\textwidth}{!}{ 
    \begin{fmfgraph*}(20,20)
      \fmfleft{i}
      \fmfright{j}
      \fmfv{d.shape=circle, d.filled=full, d.size=1mm}{i}  %
      \fmfv{d.shape=circle, d.filled=full, d.size=1mm}{j}  %
      \fmfv{d.shape=circle, d.filled=full, d.size=1mm}{k1}  %
      \fmfv{d.shape=circle, d.filled=full, d.size=1mm}{k2}  %
      \fmf{plain,tension=0.5}{k1,k1}
      \fmf{plain}{i,k1}
      \fmf{plain}{k1,k2}
      \fmf{plain}{k2,j}
    \end{fmfgraph*}
    } 
}
\end{fmffile}
+
\begin{fmffile}{DC_G74_3}
\raisebox{-0.3cm}{
    \resizebox{0.05\textwidth}{!}{ 
    \begin{fmfgraph*}(20,20)
      \fmfleft{i}
      \fmfright{j}
      \fmfv{d.shape=circle, d.filled=full, d.size=1mm}{i}  %
      \fmfv{d.shape=circle, d.filled=full, d.size=1mm}{j}  %
      \fmfv{d.shape=circle, d.filled=full, d.size=1mm}{k1}  %
      \fmfv{d.shape=circle, d.filled=full, d.size=1mm}{k2}  %
      \fmf{plain,tension=0.5}{k1,k1}
      \fmf{plain}{i,k2}
      \fmf{plain}{k2,k1}
      \fmf{plain}{k1,j}
    \end{fmfgraph*}
    } 
}
\end{fmffile}
+
\begin{fmffile}{DC_G74_4}
\raisebox{-0.3cm}{
\resizebox{0.06\textwidth}{!}{
\begin{fmfgraph*}(20,20)
  \fmftop{i}
  \fmfleft{k1}
  \fmfright{k2}
  \fmf{plain}{k1,j}
  \fmf{plain}{j,k2}
  \fmf{plain,left=1}{i,j}
  \fmf{plain,right=1}{i,j}
\fmfv{d.shape=circle,d.filled=full,d.size=1mm}{i}
\fmfv{d.shape=circle,d.filled=full,d.size=1mm}{j}
\fmfv{d.shape=circle,d.filled=full,d.size=1mm}{k1}
\fmfv{d.shape=circle,d.filled=full,d.size=1mm}{k2}
\end{fmfgraph*}
}
}
\end{fmffile}
\Bigg\}\\
&+8\frac{W(1,1,1,\bm 0)}{W(\bm 0)}
\begin{fmffile}{DC_G74_5}
\raisebox{-0.3cm}{
    \resizebox{0.05\textwidth}{!}{ 
    \begin{fmfgraph*}(20,20)
      \fmfleft{i}
      \fmfright{j}
      \fmfv{d.shape=circle, d.filled=full, d.size=1mm}{i}  %
      \fmfv{d.shape=circle, d.filled=full, d.size=1mm}{j}  %
      \fmfv{d.shape=circle, d.filled=full, d.size=1mm}{k1}  %
      \fmfv{d.shape=circle, d.filled=full, d.size=1mm}{k2}  %
      \fmfv{d.shape=circle, d.filled=full, d.size=1mm}{k3}  %
      \fmf{plain}{i,k1}
      \fmf{plain}{k1,k2}
      \fmf{plain}{k2,k3}
      \fmf{plain}{k3,j}
    \end{fmfgraph*}
    } 
}
\end{fmffile}
\end{align*}
With the diagrammatic representation for $\tilde{\bm G}_s^{(n)}$, it is straightforward to use \eqref{tilde_G_to_G} to get ${\bm G}_s^{(n)}$. Diagrammatically, this is equivalent to adding two dashed lines 
\begin{fmffile}{FU_02}
\raisebox{-0.15cm}{
    \resizebox{0.03\textwidth}{!}{ 
    \begin{fmfgraph*}(20,20)
      \fmfleft{i}
      \fmfright{j}
      \fmfv{d.shape=circle, d.filled=full, d.size=1mm}{i}  %
      \fmfv{d.shape=circle, d.filled=full, d.size=1mm}{j}  %
      \fmf{dashes}{i,j}
    \end{fmfgraph*}
    } 
}
\end{fmffile}
at the two ends of the $\tilde{\bm G}_s^{(n)}$ diagram, which cancel the neighboring solid lines.
The resulting diagrammatic expansion is equivalent to that obtained by Bender et al.~\cite{bender1979strong,bender1980strong}, and 
applying the Feynman rules described above yields the announced results Eq.~\eqref{phi_4_SCE_firstfew}. As in Ref.~\cite{bender1979strong}, 
our approach can also readily generalize to other local interactions, such as \( \phi^{2N} \) and \( (\bar{\phi}\phi)^2 \).
The main contribution of our SCE approach is that, by combining Eqs.~\eqref{combi_sum}--\eqref{feynman_string} with the general M\"obius inversion formula \eqref{Mobius_inversion_sym},
we obtain a systematic procedure for generating higher-order terms together with explicit combinatorial prefactors. 
This structure is well suited for computer-aided derivations at higher orders.

\section{Global approximation of the correlation function with Pad\'e expansions}
\label{sec:2pade_intro}
The weak- and strong-coupling expansions provide accurate approximations to the correlation function only in their respective regimes, \( g \to 0 \) and \( g \to +\infty \). To obtain a global approximation valid for all \( g \), one must analytically continue the series beyond its radius of convergence. A standard approach is to use Pad\'e approximants. In the physics context, the classical (one-point) Pad\'e approximation is normally constructed from the WCE and have been applied extensively in quantum field theory and statistical mechanics; see, e.g.,~\cite{baker1981pade,bender1991novel}. In this section, we show that combining WCE and SCE via {\em two-point Pad\'e approximations} yields a more accurate global approximation to the correlation functions of the lattice $\phi^4$ field, and achieves uniform convergence to the correct result as the expansion order increases.
We defer a concise introduction to one-point Pad\'e, two-point Pad\'e, and Borel--Pad\'e constructions used in this section to Appendix~\ref{sec:app_pade}. Here we focus on the numerical performance of these approximations in the zero-dimensional and 1D lattice \( \phi^4 \) models, and use complex-analytic arguments to  explain their convergence properties.

%
%
\subsection{Zero-dimensional $\phi^4$ field theory}
\label{sec:0d_phi4}

When the physical dimension of the lattice Hamiltonian~\eqref{lattice_phi_4} is reduced to zero, we obtain a simple schematic model: the zero-dimensional \(\phi^4\) field ~\cite{brown2015two}, which corresponds to a probability distribution over a single real variable \( q \), with probability density:
\begin{align}\label{0d_phi4_PDF}
\rho(q) \propto e^{-\frac{1}{2} m^2 q^2 - \frac{g}{4!}q^4}.
\end{align}
This model is exactly solvable and therefore provides an ideal testbed for benchmarking Pad\'e-based schemes. 
Our goal is to compute the two-point correlation function. For the zero-dimensional model~\eqref{0d_phi4_PDF}, this reduces to the second moment of \( q \):
\begin{align*}
G = \langle q^2 \rangle = \frac{1}{Z} \int_{-\infty}^{+\infty} q^2\, e^{-\frac{1}{2} m^2 q^2 - \frac{g}{4!} q^4} \, dq,
\end{align*}
where \( Z \) is the normalization constant. This integral admits an exact analytical solution \cite{brown2015two}:
\begin{align}\label{od_G_analytic}
G = \frac{4}{m^4} \rho \left[ \frac{K_{3/4}(\rho)}{K_{1/4}(\rho)} - 1 \right], \quad \text{where} \quad
\rho = \frac{3m^4}{4g},
\end{align}
where \( K_{1/4}(\cdot) \) and \( K_{3/4}(\cdot) \) are modified Bessel functions of the second kind. Throughout we fix \( m = 1 \) and study \( G = G(g) \) as a function of the coupling. The WCE and SCE for \( G(g) \) can be derived by expanding \( e^{-g q^4/4!} \) or \( e^{-\frac{1}{2} m^2 q^2} \), respectively, and then taking the ensemble average~\cite{zhu2025global}:
\begin{align}
G(g)
&=1 - \frac{g}{2} + \frac{2g^2}{3} - \frac{11g^3}{8} + \frac{34g^4}{9} + \cdots.
\label{0d_wce}\\
G(g) &= \left[\frac{2\sqrt{6}\,\Gamma\left(\frac{3}{4}\right)}{\Gamma\left(\frac{1}{4}\right)}\right] \frac{1}{\sqrt{g}} 
+ 12\left[\left(\frac{\Gamma\left(\frac{3}{4}\right)}{\Gamma\left(\frac{1}{4}\right)}\right)^2 - \frac{\Gamma\left(\frac{5}{4}\right)}{\Gamma\left(\frac{1}{4}\right)}\right] \frac{1}{g} 
+ \left\{6 \frac{\Gamma\left(\frac{7}{4}\right)}{\Gamma\left(\frac{1}{4}\right)}
- 18 \sqrt{6} \frac{\Gamma\left(\frac{5}{4}\right)\Gamma\left(\frac{3}{4}\right)}{[\Gamma\left(\frac{1}{4}\right)]^2}
+ 12 \sqrt{6} \frac{[\Gamma\left(\frac{3}{4}\right)]^3}{[\Gamma\left(\frac{1}{4}\right)]^3} \right\} \frac{1}{g^{3/2}} + \cdots.
\label{0d_sce}
\end{align}
High-order coefficients in~\eqref{0d_wce} and~\eqref{0d_sce} can be generated efficiently via recursion relations~\cite{zhu2025global}, which allows us to construct Pad\'e-type rational approximations to \(G(g)\) at high orders. Using WCE and/or SCE data, we build: (i) one-point Pad\'e approximants (1Pad\'e) from WCE and SCE which yield WCE-1Pad\'e and SCE-1Pad\'e expansion respectively, (ii) the two-point Pad\'e (2Pad\'e) interpolating the weak- and strong-coupling regimes, and (iii) the Borel--Pad\'e approximation constructed from WCE (see Appendix~\ref{sec:app_pade}). All the Pad\'e approximation results are given in terms of \(\tilde g=\sqrt g\). Figure~\ref{fig:0d_pade_compare} summarizes the numerical performance and leads to the following observations:
\begin{enumerate}
\item The truncated WCE and SCE series alone do not provide accurate approximations to \(G(\tilde g)\) over the full range of couplings. The WCE is asymptotic and diverges rapidly at large \(\tilde g\), while the SCE is accurate at large \(\tilde g\) but breaks down as \(\tilde g\to 0\).

\item Both WCE-1Pad\'e and SCE-1Pad\'e yield \emph{convergent} approximations for all \(0<\tilde g<\infty\). The convergence is not uniform: WCE-1Pad\'e slows down as \(\tilde g\) increases, whereas SCE-1Pad\'e slows down as \(\tilde g\to 0\). Overall, SCE-1Pad\'e performs better at intermediate and strong coupling.

\item Both Borel--Pad\'e and 2Pad\'e provide \emph{global} approximations to \(G(\tilde g)\), valid across the entire range \(0<\tilde g<\infty\). In general, 2Pad\'e converges faster and yields an overall more accurate results than Borel--Pad\'e across the full \(\tilde g\)-range in our tests.
\end{enumerate}

A detailed comparison of convergence rates for SCE-1Pad\'e, Borel--Pad\'e, and 2Pad\'e at representative coupling values is shown in Fig.~\ref{fig:0d_pade_convergence_rate}. From the figure, we see that all three approaches exhibit exponential convergence with order. Borel--Pad\'e (built from WCE) converges fastest at small \(\tilde g\), while SCE-1Pad\'e converges fastest at large \(\tilde g\), indicating that each rational scheme retains analytic information from the series used in its construction. In comparison, the convergence rate of 2Pad\'e approximation is \emph{uniform} across the full range of \(\tilde g\), consistent with its role as an interpolant between weak and strong coupling.

\begin{figure}[t]
\centering
\includegraphics[width=18cm]{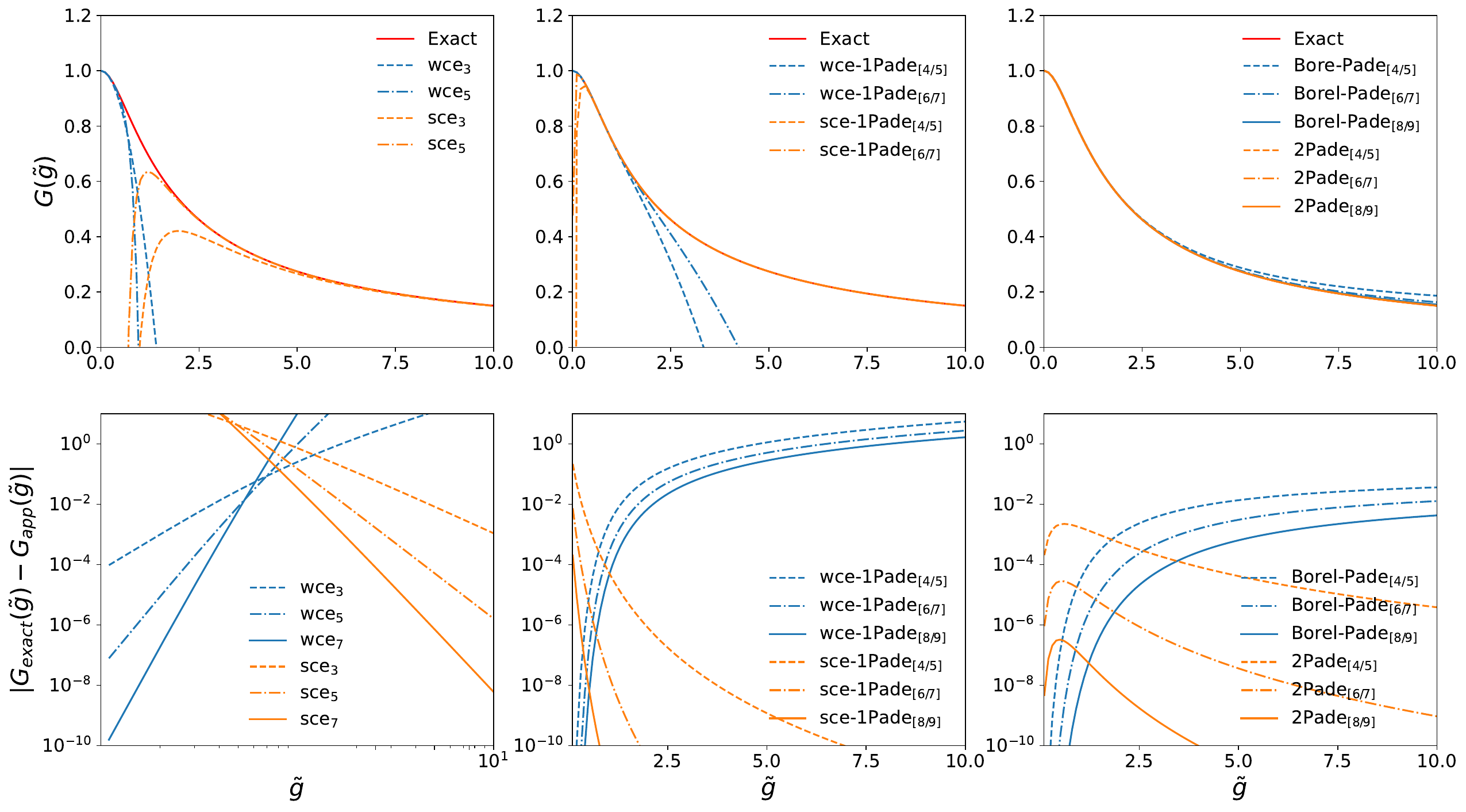}
\caption{Different approximations of \( G(\tilde{g}) \) for the zero-dimensional $\phi^4$ field theory, where \( \tilde{g} = \sqrt{g} \) is used. All the subscripts correspond to the order of the approximation. The second panel shows the approximation error relative to the exact result \eqref{od_G_analytic}. Note that the error plots for the WCE and SCE expansions are shown on a log-log scale, which clearly illustrates the divergence of the SCE at small values of \( g \).
}
\label{fig:0d_pade_compare}
\end{figure}

\begin{figure}[t]
\centering
\includegraphics[width=18cm]{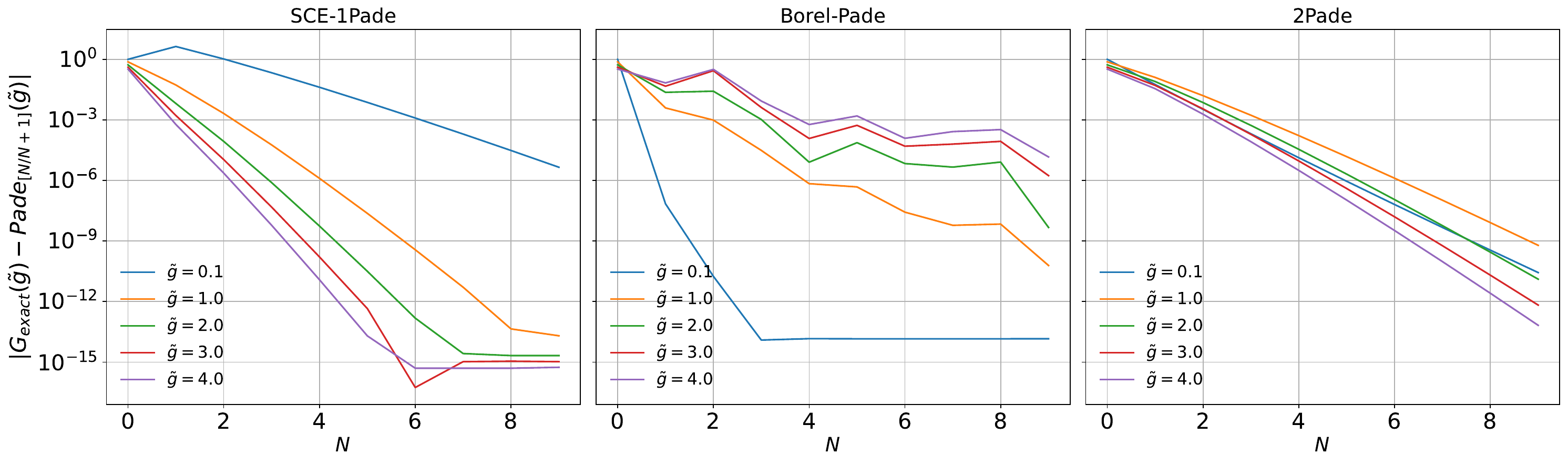}
\caption{Convergence rate comparison between SCE-1Pad\'e, Borel-Pad\'e, and 2Pad\'e expansion for zero-dimensional $\phi^4$ field theory at selected interaction strength $\tilde g=\sqrt{g}$. 
}
\label{fig:0d_pade_convergence_rate}
\end{figure}
\subsection{One-dimensional lattice $\phi^4$ field theory}
\label{sec:1d_phi4}
Next, we apply the Pad\'e expansion techniques to the one-dimensional lattice \( \phi^4 \) field theory and compute the two-point correlation function \( G_{ij}(g) \) defined in~\eqref{two-point_G}. In this setting, the weak-coupling expansion (WCE) can be derived by standard perturbation theory and Feynman-diagram techniques~\cite{parisi1988statistical}. In this work, we use WCE coefficients up to third order; the corresponding diagrams are given in Ref.~\cite{brown2015two}. The strong-coupling expansion (SCE) up to third order was obtained in the previous sections, see~\eqref{phi_4_SCE_firstfew}. Since \( G_{ij}(g) \) is matrix-valued, the Pad\'e expansion is applied element-wise to construct the approximation. As the one-dimensional model is no longer exactly solvable, we validate the resulting approximations by comparing with Monte Carlo estimates obtained using a Langevin sampler.

The numerical results are summarized in Fig.~\ref{fig:1d_pade}. Since we only computed the WCE and SCE up to third order \footnote{The WCE can be computed to much higher orders using advanced symbolic and combinatorial techniques~\cite{kleinert2001critical,serone2018lambdaphi4}.}, the truncated WCE and SCE series, as well as one-point Pad\'e/Borel--Pad\'e approximations built from one-sided information, perform poorly. In contrast, the 2Pad\'e approximation provides a uniformly accurate result across the full range of interaction strengths shown. This behavior is expected: combining WCE and SCE information through third order provides enough constraints to construct a $[3/4]$-th order 2Pad\'e approximant. By comparison, achieving a third-order Pad\'e approximation from only one-sided information (either WCE or SCE) would require at least seven expansion coefficients, i.e., WCE or SCE through sixth order. Such calculations are highly nontrivial, since the number of Feynman diagrams grows factorially with order~\cite{dorigoni2019introduction}. This example clearly illustrates another numerical advantage of 2Pad\'e: by combining \(N\)-th order WCE and SCE data, it yields an accurate \(N\)-th order rational approximation that would otherwise require a \((2N+1)\)-st order one-sided construction. In our work~\cite{zhu2025global}, we also presented additional numerical results for 1D lattice $\phi^4$ theory with various lattice sizes, $t, \mu$ and for the off-diagonal component of $G_{ij}(g)$. These results confirm that the 2Pad\'e expansion yields convergent approximations to \( G_{ij}(g) \), and that its convergence properties---such as the shape of the error curve and the convergence rate---are consistent with those observed in the zero-dimensional $\phi^4$ field theory.

\begin{figure}[t]
\centering
\includegraphics[width=14cm]{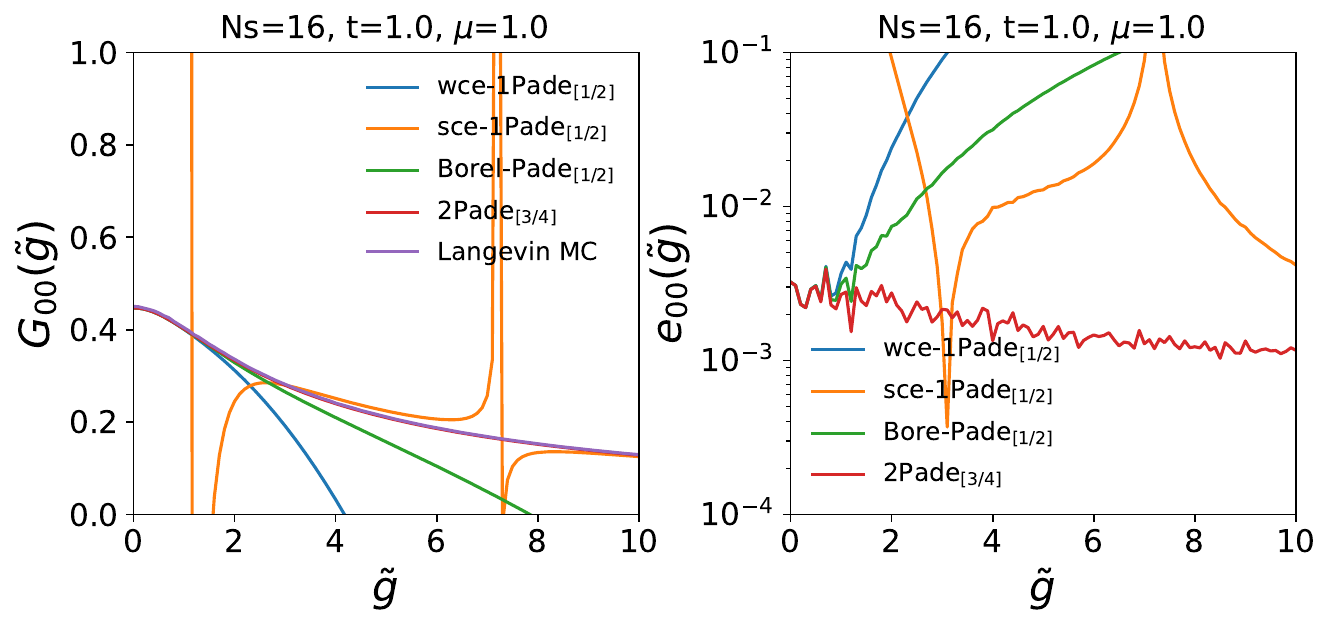}
\caption{
(\textbf{Left}) Different Pad\'e approximation for $G_{00}(\tilde g)$ of the 1D lattice $\phi^4$ field and (\textbf{Right}) their corresponding approximation error $e_{00}(\tilde g)=|G^{MC}_{00}(\tilde g)-G^{app}_{00}(\tilde g)|$ . All the subscripts correspond to the order of the approximation. Note that using only WCE or SCE up to the third order, the highest order 1Pad\'e or Borel-Pad\'e expansion one could get is $[1/2]$, which is what we displayed here. On the contrary, 2Pad\'e combines WCE and SCE hence we can get expansion up to the order $[3/4]$. The simulated $\phi^4$ field has lattice site $N_s=16$ with $t=\mu=1.0$. The Langevin MC results are based on \( 10^8 \) samples, yielding a numerical accuracy of approximately \( O(10^{-4}) \). Due to this limited accuracy, the error curves for 2Pad\'e is less accurate. By the analysis of our work \cite{zhu2025global}, the numerical accuracy of 2Pad\'e$_{[3/4]}$ is expected to be slightly smaller than $10^{-3}$ for all $\tilde g$.
}
\label{fig:1d_pade}
\end{figure}
\subsection{Convergence Analysis}
The numerical results in the previous sections clearly demonstrate that the 2Pad\'e expansion converges to the exact correlation function of the lattice \(\phi^4\) theory. More broadly, the (often surprising) convergence of one-point Pad\'e (1Pad\'e) approximations in capturing thermodynamic quantities and correlation functions has been observed in a variety of many-body systems, including the Ising model~\cite{baker1981continuous,george2012quantitative}. In some special cases, such convergence can even be established rigorously~\cite{baker1981continuous,parisi1988statistical}. However, a general and fully rigorous convergence theory for Pad\'e-type approximations in interacting many-body settings remains out of reach due to the well-known difficulties in mathematical analysis~\cite{stahl1997convergence,buslaev2013convergence}. Accordingly, the goal of this section is {\em not} to provide a complete mathematical proof of the observed convergence, but rather to offer a heuristic explanation based on the analytic structure of the correlation functions. Some of the arguments were introduced in our previous work~\cite{zhu2025global}; here we provide additional numerical evidence and further discussion. In particular, we focus on two aspects: (i) how the SCE resolves the singularity and leads to the local convergence at strong coupling, and (ii) the analytic structure of \(G(g)\) in the complex plane and its implications for the convergence of 2Pad\'e approximations.


\subsubsection{Resolution of singularity by strong-coupling expansion}
\begin{figure}[t]
\centering
\includegraphics[width=18cm]{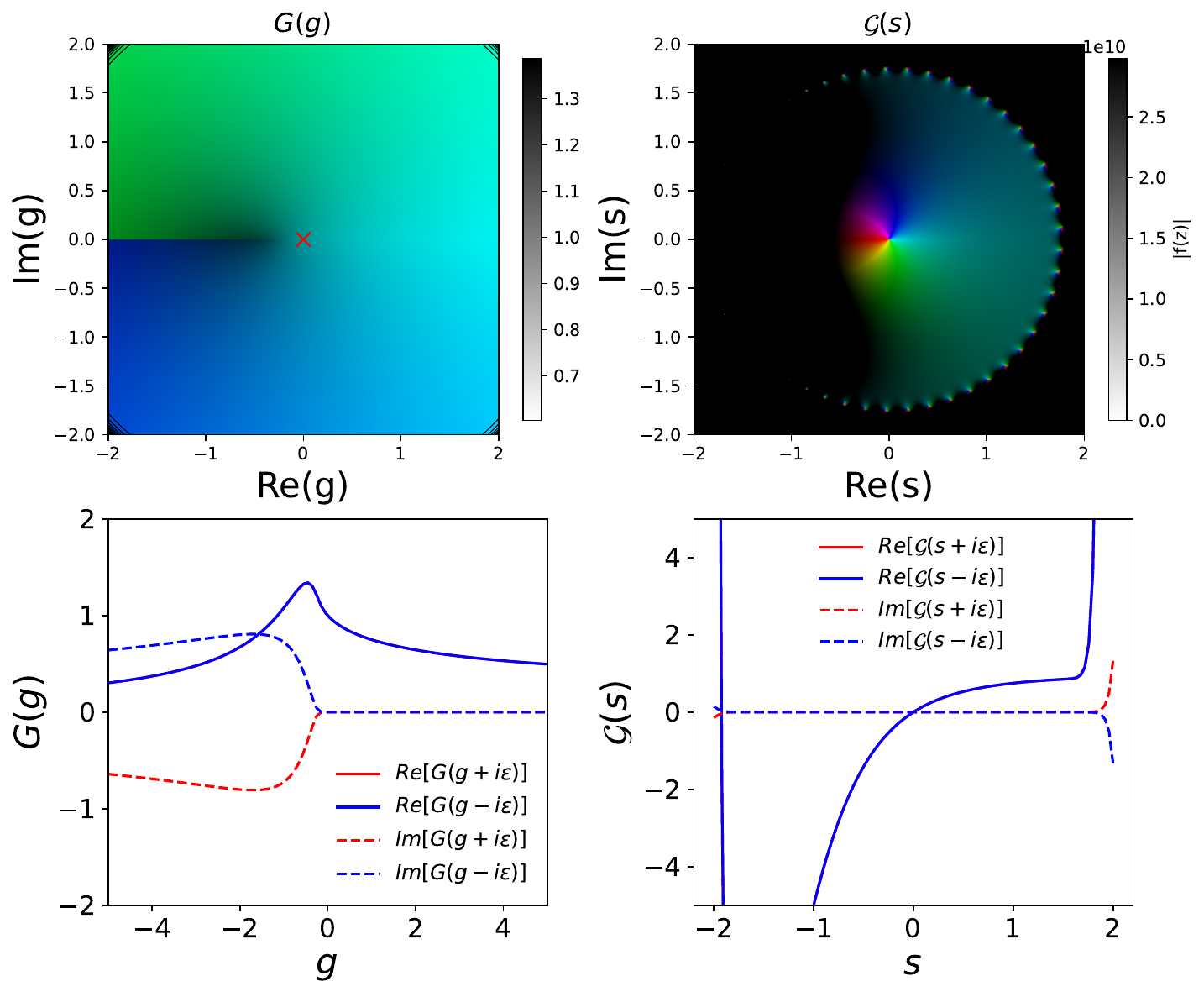}
\caption{\textbf{Top row:} Domain coloring plots of \( G(g) \) and \( \mathcal{G}(s) \) in the complex plane. The grayness indicates the modulus of the function, while the hue represents the complex argument \( \theta \). The branch point at \( g = 0 \) is fully resolved in the \( s \)-plane, where \( \mathcal{G}(s) \) is analytic at \( s = 0 \). \textbf{Bottom row:} Plots of \( G(g \pm i\epsilon) \) and \( \mathcal{G}(s \pm i\epsilon) \) with a small imaginary shift \( \epsilon = 10^{-4} \). The branch-point singularity of \( G(g) \) at \( g = 0 \) is clearly visible, in contrast to the regular behavior of \( \mathcal{G}(s) \). We emphasize that the plot of \( \mathcal{G}(s) \) outside the moon-shaped convergence disk is inaccurate, as it is generated from the truncated 50th-order power series expansion~\eqref{0d_sce_remove}. As we discuss below, the correct continuation of \( \mathcal{G}(s) \) beyond this disk is well captured by Pad\'e approximation (see Fig.~\ref{fig:domain_coloring_pade}).
}
\label{fig:domain_coloring_sce}
\end{figure}
For the \(\phi^4\) field theory, a first striking feature is the different convergence behavior of the WCE and SCE. We use the zero-dimensional example to explain this mechanism. Consider the SCE around \(g=\infty\). Introducing \(s=1/\sqrt{g}\) and analytically continuing \(s\) to the extended complex plane \(\eC=\C\cup\{\infty\}\), the SCE~\eqref{0d_sce} becomes a power series that defines a germ of a new function \( \mathcal{G}(s) \) at \(s=0\):
\begin{align}
\mathcal{G}(s) = \left[\frac{2\sqrt{6}\,\Gamma\left(\frac{3}{4}\right)}{\Gamma\left(\frac{1}{4}\right)}\right] s
+ 12\left[\left(\frac{\Gamma\left(\frac{3}{4}\right)}{\Gamma\left(\frac{1}{4}\right)}\right)^2 - \frac{\Gamma\left(\frac{5}{4}\right)}{\Gamma\left(\frac{1}{4}\right)}\right] s^2
+ \cdots.
\label{0d_sce_remove}
\end{align}
In \( \eC \), the mapping \( s = 1/\sqrt{g} \) is multivalued.
Thus \( \mathcal{G}(s) \) defined as \eqref{0d_sce_remove} differs from the analytic continuation of the exact solution
\eqref{od_G_analytic} around \( g = \infty \).
We have not found a closed-form expression for \( \mathcal{G}(s) \) in terms of
special functions, although it might exist.
Nevertheless, by numerically analyzing the decay rate of the coefficients in
\eqref{0d_sce_remove} up to order 50, we observe a geometric decay with approximate rate
\( \alpha = 1.8 \).
This suggests that \( \mathcal{G}(s) \) is analytic within the disk
\( |s| \leq 1/\alpha \), which explains why the SCE converges for \( g \gtrsim 3.24 \).
A similar analytic continuation was performed for the WCE by Brown et al.~\cite{brown2015two},
who showed that the WCE series exhibits a branch-point singularity at \( g = 0 \) in the
complex plane.
This singularity fundamentally explains the divergence of the WCE around \( g = 0 \).
By switching to the SCE, we effectively ``resolve'' the singularity at the origin by
reparametrizing the problem in terms of \( s \).
A visualization of \( \mathcal{G}(s) \) and the analytic continuation of the WCE in the
complex plane is shown in Fig.~\ref{fig:domain_coloring_sce}.
For the 1D (and more generally \(Nd\)) lattice \(\phi^4\) model, we expect an analogous
mechanism at the level of each matrix element: the strong-coupling expansion defines a
function \(\mathcal{G}_{ij}(s)\) that is analytic near \(s=0\), thereby removing the
branch-point obstruction at \(g=0\) seen in the WCE variable. 


%
%
\subsubsection{Analytic continuation by Pad\'e expansion}
\begin{figure}[t]
\centering
\includegraphics[width=18cm]{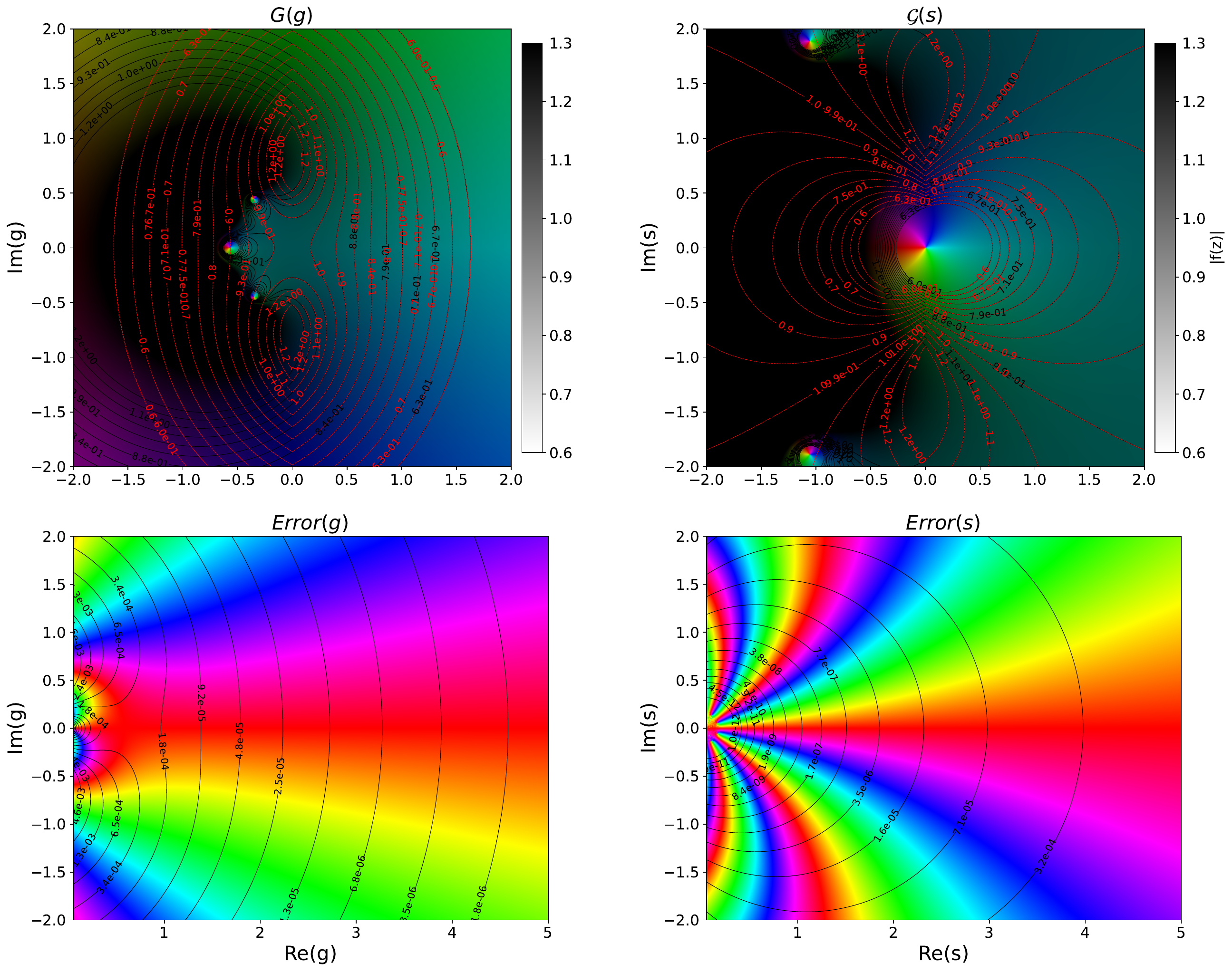}
\caption{\textbf{(Top Left)} Comparison between the analytic continuation (AC) of the exact solution \( G(g) \) given in Eq.~\eqref{od_G_analytic} and that of its 4th-order 2Pad\'e approximation \( G_{\text{Pad\'e}}(g) \) in the complex plane. The red contour lines represent the modulus \( |G(g)| \), and the black contour lines show \( |G_{\text{Pad\'e}}(g)| \). The near-perfect overlap in the right half-plane (\( \text{Re}[g] > 0 \)) demonstrates good agreement between the two.  
\textbf{(Bottom Left)} Plot of the approximation error \( G_{\text{Pad\'e}}(g) - G(g) \), highlighting the accuracy of the 2Pad\'e approximation for \( \text{Re}[g] \geq 0 \). \textbf{(Top Right)} The same comparison between \( \G(s) \) and that of its 5th-order SCE-1Pad\'e approximation \( \G_{\text{Pad\'e}}(s) \) in the complex plane.
\textbf{(Bottom Right)} Plot of the approximation error \( \G_{\text{Pad\'e}}(s) - \G(s) \), highlighting the accuracy of the SCE-1Pad\'e approximation for \( \text{Re}[s] \geq 0 \).
}
\label{fig:domain_coloring_pade}
\end{figure}
The analyticity of \( \G(s) \) in a neighborhood of \( s=0 \) implies that one can perform analytic continuation (AC) from this germ to obtain a function analytic on a larger domain of the complex plane. In practice, Pad\'e approximation provides a convenient numerical realization of such an AC procedure. To see this clearly, Figure~\ref{fig:domain_coloring_pade} compares the AC of the exact solutions \(G(g)\) and \(\G(s)\) with two convergent Pad\'e approximants constructed in Section~\ref{sec:0d_phi4}, namely SCE-1Pad\'e and 2Pad\'e, in the complex plane. Even at relatively low order, the Pad\'e approximants closely track the exact functions in the right half-plane, where \(G(g)\) and \(\G(s)\) are expected to be analytic. Together with the resolution-of-singularity mechanism provided by the SCE, this suggests the following explanation for the observed convergence of SCE-1Pad\'e and 2Pad\'e:
\begin{itemize}
\item \textbf{Analytic germ (SCE).} The SCE supplies an analytic germ for the target observable by reparametrizing the problem in terms of \(s=1/\sqrt g\). In this variable, the branch-point obstruction at \(g=0\) is removed and \(\G(s)\) is analytic near \(s=0\).
\item \textbf{Analytic continuation (Pad\'e).} Pad\'e approximation then performs a numerical AC from this germ \cite{baker1981pade}, producing rational approximants that can represent \(\G(s)\) (and hence \(G(g)\)) on a much larger region than the original power series.
\item \textbf{Analyticity for finite lattices.} For finite lattice \(\phi^4\) models, the Lee--Yang theory~\cite{lee1952statistical,yang1952statistical} suggests that true thermodynamic non-analyticities are absent at finite size. Accordingly, for any fixed \(m\), \(G(g)\) is expected to be analytic in a $\delta$-tube encircling the positive real axis. By uniqueness of analytic continuation, any convergent Pad\'e sequence that matches the germ must converge to the same analytic function, i.e., the exact \(G(g)\).
\item \textbf{Extrapolation vs.\ interpolation.} SCE-1Pad\'e illustrates that AC from a single germ can already yield convergent approximations, but the convergence rate need not be uniform in \(g\) because the continuation is essentially \emph{extrapolative}. By incorporating information from both weak and strong coupling, 2Pad\'e makes the continuation more \emph{interpolative}, which empirically leads to more robust and nearly uniform convergence over \(\mathrm{Re}\,g>0\).
\end{itemize}
In the thermodynamic limit genuine non-analyticities may appear due to phase transitions. Nevertheless, Pad\'e-type approximants can still provide accurate approximations for finite (but sufficiently large) systems. In practice, one may apply Pad\'e approximants to finite systems with sites $N$, and then gradually take the thermodynamic limit $N\rightarrow+\infty$ until the finite-size effect becomes negligible. 

The above convergence analysis should be applicable to $N$-dimensional lattice $\phi^4 $ theory for any matrix element \( G_{ij}(g) \), and it can be expected to hold for a broader class of many-body systems such as the Hubbard model \cite{zhu2025global}. Yet, we emphasize again that the above arguments are heuristic and not mathematically rigorous. In particular, the analyticity of correlation function (in our case, \( \G(s) \)) in a neighborhood of the positive real axis is a widely held belief but has not been rigorously established for general field theory models except 2D Ising and several other models. Moreover, the uniqueness of the analytic continuation merely implies that when Pad\'e-type approximants converge, it will converge to the right result (under the analyticity assumption, see Faithfulness theorem in \cite{baker1981continuous}). But the rigorus proof of the Pad\'e convergence is rather technical \cite{stahl1997convergence,buslaev2013convergence} and relies on assumptions on the underlying functions which are hard to check for practical applications.

\subsubsection{Comparison with Borel resummation}

The results obtained for the $\phi^4$ model suggest that the SCE and 2Pad\'e expansions provide a promising route to non-perturbative approximations in strongly interacting systems. It is therefore instructive to compare our approach with Borel resummation~\cite{parisi1988statistical} and with more recent developments in resurgence theory~\cite{dorigoni2019introduction,marino2014lectures}. In the Borel framework, the divergent perturbative series is first Borel transformed, yielding a function $\mathcal{B}[\phi](\zeta)$ whose Taylor series converges near $\zeta=0$ in the Borel plane. If $\mathcal{B}[\phi](\zeta)$ is resurgent, one can analytically continue it—either exactly or via Pad\'e approximants—through the complex $\zeta$-plane and then recover the original function by an inverse Laplace transform.

In contrast, our approach resolves the singularity by introducing a new expansion in a variable \(s=s(z)\) that admits an analytic germ at \(s=0\). One can then apply 1Pad\'e or 2Pad\'e approximants to analytically continue this germ and obtain a global approximation to the target function \(G(s)\). Assuming that \(G(s)\) is at least analytic in a neighborhood of positive real axis in the complex plane, uniqueness of analytic continuation implies that whenever the Pad\'e approximants converge, they must converge to the correct function. In this work the alternative expansion is provided by the SCE with \(s=1/\sqrt{g}\). We expect that other asymptotic expansions---for example, a large-\(N\) expansion---could play an analogous role, and that the 2Pad\'e framework may be useful more broadly for interpolating between distinct asymptotic regimes.

From a theoretical viewpoint, the 2Pad\'e approach requires additional effort because it relies on deriving an alternative expansion with an analytic germ, a task that is often technical and system dependent. However, once such an expansion is available, its practical value can be substantial. As discussed in Section~\ref{sec:1d_phi4}, one can combine low-order coefficients from the original asymptotic series (WCE) with those from the alternative expansion (SCE) to construct a relatively high-order 2Pad\'e approximant. Achieving a comparable accuracy with a one-sided method (1Pad\'e or Borel--Pad\'e) would typically require many more perturbative coefficients. For realistic statistical and quantum field theories, this can dramatically reduce the number of Feynman diagrams that must be computed, and also enabling more accurate approximations at significantly lower cost.

\section{Conclusion}
\label{sec:Conclusion}
In this work, we develop a practical route to global, non-perturbative approximations for correlation functions of lattice \(\phi^4\) theory by combining two complementary ingredients. First, we provided a new combinatorial derivation of the SCE that yields coefficients in a form amenable to systematic generation. Second, we introduced two-point Pad\'e (2Pad\'e) approximations that match weak- and strong-coupling expansion series and produce accurate and convergent approximations of the correlation function across the full range of coupling strength. The observed numerical convergence of the Pad\'e approximants suggests that for \emph{finite} lattice systems where the correlation functions are expected to be analytic, Pad\'e-type approximants provides a numerical analytic continuation of perturbative series, and can converge to the underlying correlation function by uniqueness of analytic continuation. Among the different Pad\'e approximation schemes we tested, 2Pad\'e offers a clear numerical advantage: by incorporating information from both asymptotic regimes it yields a more robust and nearly uniform convergence than constructions based solely on one-sided input (WCE-1Pad\'e/Borel--Pad\'e or SCE-1Pad\'e). Moreover, constructing a $N$-th order 2Pad\'e approximant requires only $N$ coefficients from each expansion, whereas a one-sided method would typically require $2N+1$ or more coefficients to achieve comparable accuracy. This dramatically reduce the computational cost of enumerating Feynman diagrams for obtaining accurate approximations in realistic field theories.

Several directions remain open. Computationally, extending the SCE and WCE to higher orders is challenging due to the factorial growth of number of diagrams, but it is precisely in this regime that the combinatorial SCE and two-point matching can be most advantageous. Theoretically, a fully rigorous convergence theory for Pad\'e-type approximants in interacting many-body systems is still missing , and it would be interesting to see how 2Pad\'e approximants behave near the thermodynamic limit where genuine non-analyticities start to arise. More broadly, the 2Pad\'e philosophy applies whenever two complementary asymptotic regimes are available, suggesting applications beyond the \(\phi^4\) setting, e.g., to general field theory models with large-\(N\), high-temperature/low-temperature, or weak/strong-coupling dual expansions.

\section{Acknowledgement} 
This material is based upon work supported by the U.S. Department of Energy, Office of Science, Office of Advanced Scientific Computing Research and Office of Basic Energy Sciences, Scientific Discovery through Advanced Computing (SciDAC) program under Contract No. DE-AC02-05CH11231. This work is also supported by the Center for Computational Study of Excited-State Phenomena in Energy Materials (C2SEPEM) at the Lawrence Berkeley National Laboratory, which is funded by the U.S. Department of Energy, Office of Science, Basic Energy Sciences, Materials Sciences and Engineering Division, under Contract No. DE-AC02-05CH11231, as part of the Computational Materials Sciences Program. This research used resources of the National Energy Research Scientific Computing Center, a DOE Office of Science User Facility supported by the Office of Science of the U.S. Department of Energy under Contract No. DE-AC02-05CH11231 using ASCR-ERCAP-m1027. Y. Zhu acknowledges valuable discussions with Zhen Huang, Emanuel Gull and Yang Yu.   



\appendix
\section{Möbius inversion formula}\label{sec:app1_mobius}
This appendix records a convenient identity for rewriting \emph{exclusion} sums (all indices distinct) in terms of \emph{inclusion} sums (unrestricted sums with possible index coincidences) in diagrammatic expansions. This identity is useful because the inclusion sum is  compatible with the standard Feynman rules which can typically be evaluated efficiently (for example, as tensor contractions), while exclusion constraints such as \(k_1\neq k_2\neq\cdots\) are cumbersome to implement directly. Fix an integer \(d\ge 1\) and define
\[
V=\{1,\ldots,N\}^d,\qquad
W=\{(k_1,\ldots,k_d)\in V:\ k_a\neq k_b\ \text{for all }a\neq b\}.
\]
Let \(S_d\) be the permutation group on \(\{1,\ldots,d\}\). For \(v=(v_1,\ldots,v_d)\in V\) and \(\pi\in S_d\), let \(\pi(v)\) be the vector obtained by permuting the \emph{positions},
\(\pi(v)=(v_{\pi^{-1}(1)},\ldots,v_{\pi^{-1}(d)})\).
For a function \(f:V\to \mathcal{A}\) valued in a vector space or algebra \(\mathcal{A}\) (possibly noncommutative), define the stabilizer \(\mathrm{Stab}(v)=\{\pi\in S_d:\ \pi(v)=v\}\). Then the exclusion sum can be written as an alternating sum over stabilizers:
\begin{align}\label{Mobius_decompose}
\sum_{w \in W} f(w)
= \sum_{v \in V}\Bigg(\sum_{\pi \in \mathrm{Stab}(v)} (-1)^{\text{sign}(\pi)}\Bigg) f(v)
= \sum_{\substack{v \in V,\, \pi \in S_d \\ \pi(v) = v}} (-1)^{\text{sign}(\pi)} f(v).
\end{align}
To see this, note that if \(v\) has at least one repeated index, then \(\mathrm{Stab}(v)\) contains a transposition, so even and odd permutations in \(\mathrm{Stab}(v)\) cancel and the inner sum vanishes. If \(v\in W\) has all distinct indices, then \(\mathrm{Stab}(v)=\{\mathrm{id}\}\) and the inner sum equals \(1\). To rewrite \eqref{Mobius_decompose} in terms of inclusion sums, one groups vectors \(v\in V\) by their coincidence pattern. Let \([d]=\{1,\ldots,d\}\), and let \(P\) be a partition of \([d]\) specifying which positions are forced to be equal. Define
\[
V_P=\{v\in V:\ v_a=v_b\ \text{whenever }a,b\ \text{belong to the same block of }P\}.
\]
If the block sizes of \(P\) are \(\lambda_1,\lambda_2,\ldots\) (so \(\sum_i \lambda_i=d\)), then the alternating stabilizer sum depends only on \(P\) and equals the Möbius function of the partition lattice, \(\mu(P)=(-1)^{d-|P|}\prod_i (\lambda_i-1)!\). Consequently,
\begin{equation}\label{Mobius_inversion}
\sum_{w \in W} f(w)
= \sum_{P} \mu(P)\sum_{v \in V_P} f(v)
= \sum_P (-1)^{d - |P|}\left(\prod_i (\lambda_i - 1)!\right)\sum_{v \in V_P} f(v),
\end{equation}
where the sum runs over all partitions \(P\) of \([d]\) and \(|P|\) is the number of blocks. When \(f\) is symmetric under permutations of its arguments, the terms in \eqref{Mobius_inversion} can be grouped by \(\lambda\), the integer partition of \(d\) given by the block sizes. Let \(V_\lambda\) be the set of vectors where the first \(\lambda_1\) entries are equal, the next \(\lambda_2\) entries are equal, and so on, and let \(m_1,m_2,\ldots\) denote the multiplicities of block sizes $1,2, \cdots$ in \(\lambda\). Then
\begin{equation}\label{Mobius_inversion_sym}
\begin{aligned}
\sum_{w \in W} f(w)
&= \sum_\lambda (-1)^{d - \text{len}(\lambda)} \cdot \left( \prod_i (\lambda_i - 1)! \right) \cdot (\# \text{partitions with shape } \lambda) \cdot \sum_{v \in V_\lambda} f(v)\\
&= \sum_\lambda (-1)^{d - \text{len}(\lambda)} \cdot \left( \prod_i (\lambda_i - 1)! \right) \cdot \binom{d}{\lambda_1,\lambda_2\cdots}\frac{1}{m_1! m_2! \cdots} \cdot \sum_{v \in V_\lambda} f(v)
\end{aligned}
\end{equation}
Using \eqref{Mobius_inversion} (or \eqref{Mobius_inversion_sym} when applicable), one can systematically rewrite exclusion sums such as \(\sum_{k_1\neq k_2\neq k_3}\) as linear combinations of inclusion sums such as \(\sum_{k_1k_2k_3}\) and \(\sum_{k_1k_2}\), with explicit combinatorial prefactors. For example, when \(d=3\) one obtains
\[
\sum_{i\neq j\neq k} f(i,j,k)
= \sum_{i,j,k} f(i,j,k)
 - \sum_{i,k} f(i,i,k) - \sum_{i,j} f(i,j,i) - \sum_{i,j} f(i,j,j)
 + 2\sum_i f(i,i,i),
\]
corresponding to the five partitions of \(\{1,2,3\}\) with combinatorial coefficients \(1,-1,-1,-1,2\).

\section{Pad\'e-type approximations}\label{sec:app_pade}
This appendix summarizes the Pad\'e-type approximations used in Section~\ref{sec:2pade_intro}. The common theme is to approximate a target function by a rational ansatz and to fix its coefficients by matching asymptotic information. Such rational approximations often provide an efficient analytic continuation of truncated series and can capture poles, and branch-cut structure behavior that are inaccessible to low-order polynomials. Below we outline the three approximations used in this work.

\paragraph{One-point Pad\'e approximants (1Pad\'e).}
Given a (formal) power series expansion of a function $f(x)$ around a single point, for example
\begin{align}\label{eq:1pade_wce}
f(x)=a_0+a_1 x + a_2 x^2+\cdots \qquad (x\to 0),
\end{align}
the one-point Pad\'e approximant \(P_{[L/M]}(x)\) is a rational function whose Taylor expansion agrees with this power series up to the highest possible order:
\begin{equation}
P_{[L/M]}(x)=\frac{\sum_{\ell=0}^{L} A_\ell x^\ell}{1+\sum_{m=1}^{M} B_m x^m},
\qquad
P_{[L/M]}(x)-f(x)=O(x^{L+M+1}).
\end{equation}
In practice, the coefficients \(A_\ell\) and \(B_m\) are determined by a linear system obtained by expanding the rational ansatz about the expansion point and equating coefficients up to order \(L+M\). In this work, we use 1Pad\'e either from WCE data (around \(g=0\)) or from SCE data expressed in a small parameter such as \(s=1/\sqrt{g}\) (around \(g=+\infty\)). Heuristically, 1Pad\'e can be viewed as a rational extrapolation of a local series: it often converges beyond the disk of convergence of the original Taylor series when the target function is meromorphic or admits a Stieltjes-type representation; see, e.g.,~\cite{bender2013advanced,baker1981pade}.

\paragraph{Two-point Pad\'e approximants (2Pad\'e).}
Unlike 1Pad\'e, which uses a single series expansion, the two-point Pad\'e approach assumes knowledge of two series expansions for a scalar function \( f(x) \): one around \( x = 0 \) (WCE), and another around \( x = +\infty \) (SCE),
\begin{align}
f(x) &= a_0 + a_1 x + a_2 x^2 + \cdots, \qquad x \to 0 , \label{0d_pade_wce} \\
f(x) &= \frac{b_1}{x} + \frac{b_2}{x^2} + \cdots, \qquad x \to +\infty. \label{0d_pade_sce}
\end{align}
The \( N \)-th order 2Pad\'e approximation, denoted \( P_{[N/N+1]}(x) \), approximates \( f(x) \) by a rational function
\begin{equation}\label{2pade_ansatz}
P_{[N/N+1]}(x) = \frac{A_0 + A_1 x + \cdots + A_N x^N}{1 + B_1 x + \cdots + B_{N+1} x^{N+1}},
\end{equation}
where the Pad\'e coefficients \( \{A_i\}_{i=0}^{N} \) and \( \{B_j\}_{j=1}^{N+1} \) are determined by matching the expansion of~\eqref{2pade_ansatz} with~\eqref{0d_pade_wce} as \( x \to 0 \) up to order \( O(x^{N}) \), and with~\eqref{0d_pade_sce} as \( x \to +\infty \) up to \( O(x^{-N}) \). Operationally, this is again a linear system for the unknown rational coefficients, but now the constraints come from \emph{both} ends of the domain. The resulting approximant is designed to interpolate between the weak- and strong-coupling regimes while remaining a single closed-form function of \(x\). The explicit linear systems and solvability conditions can be found in~\cite{zhu2025global,roy2009global}. To apply the 2Pad\'e approximation to the \( \phi^4 \) model, we simply reshape the WCE and SCE expansion series in term of $\tilde g=\sqrt{g}$, and then apply the 2Pad\'e approximation to the resulting series.

\paragraph{Borel--Pad\'e approximation.}
When a perturbative expansion \eqref{eq:1pade_wce} is divergent but asymptotic (as is common for the WCE of bosonic fields), one can first form its Borel transform,
\begin{equation}
\mathcal{B}f(\zeta)=\sum_{n\ge 0}\frac{a_n}{n!}\,\zeta^n,
\end{equation}
which often has a nonzero radius of convergence and the factorial growth of coefficients is converted into finite-distance singularities in the Borel plane. One then applies a one-point Pad\'e approximant to \( \mathcal{B}f(\zeta) \) to perform analytic continuation in the Borel plane, and finally reconstructs \(f\) via a Laplace-type integral along a suitable contour (when it exists),
\begin{equation}
f(x)=\int_{0}^{\infty} e^{-\zeta}\, \mathcal{B}f(\zeta x)\, d\zeta,
\end{equation}
In Section~\ref{sec:0d_phi4}, we follow this standard Borel--Pad\'e strategy as in~\cite{brown2015two,dorigoni2019introduction}.
\bibliography{main_JPA}
\end{document}